\documentclass[12pt,a4paper]{article}
\pdfoutput=1
  \usepackage{jheppub}

\numberwithin{equation}{section}

\def\qb#1{\left[ #1\right]\hspace{-1pt} !}
\def\qn#1{\left[ #1\right]_b}
\def\qnP#1{\left[ #1\right]^{\prime}_q}
\def\bea{\begin{eqnarray*}}
\def\eea{\end{eqnarray*}}
\def\Uqosp{U$_q$(osp(1$|$2))}
\def\Uqtwo{U$_q$(sl(2))}
\def\qbp#1{\left[ #1\right]_b}
\def\qbm#1{\left[ #1\right]_{\frac{1}{b}}}

\title{The universal Racah-Wigner symbol for U$_q$(osp(1$|$2))}

\author[a]{Michal Pawelkiewicz}
\author[a]{Volker Schomerus}
\author[a,b]{and Paulina Suchanek}

\affiliation[a]{DESY Theory Group, DESY Hamburg\\
                                                  Notkestrasse 85, D-22603 Hamburg, Germany}
\affiliation[b]{Institute for Theoretical Physics, University of Wroc{\l}aw, \\
 pl.~M.~Borna 9, 50-204 Wroc{\l}aw, Poland}

\emailAdd{ michal.pawelkiewicz@desy.de, volker.schomerus@desy.de, paulina.suchanek@ift.uni.wroc.pl}

\abstract{
We propose a new and elegant formula for the Racah-Wigner symbol of
self-dual continuous series of representations of \Uqosp. It
describes the entire fusing matrix for both NS and R sector of N=1
supersymmetric Liouville field theory. In the NS sector, our formula
is related to an expression derived in \cite{Hadasz:2013bwa}. Through
analytic continuation in the spin variables, our universal expression
reproduces known formulas for the Racah-Wigner coefficients of finite
dimensional representations.}

\arxivnumber{1307.6866 }

\begin{document}

\maketitle
 \flushbottom



\setcounter{equation}{0}

\section{Introduction}

The Racah-Wigner coefficients of Lie (super)algebras and their deformations
play an important role in modern mathematical physics. Up to some normalization
dependent prefactors, they coincide with the so-called fusing matrix of
2-dimensional Wess-Zumino-Novikov-Witten (WZNW) models and hence feature
very prominently in the conformal bootstrap of these models and many
descendants thereof. In fact, they do not only provide the coefficients
in the bootstrap equations but also furnish some of their famous solutions
e.g. for the bulk and boundary operator product coefficients. This dual
purpose of the Racah-Wigner coefficients is based on a number of identities
they satisfy, most importantly the well-known pentagon equation. The same
identities are also exploited in the construction of state-sum models for
topological 3-manifold invariants. These provide another important area
in which Racah-Wigner symbols appear.

Recently, two of the authors and Leszek Hadasz constructed the
Racah-Wigner symbol for a series of self-dual representations of
\Uqosp\ \cite{Hadasz:2013bwa} for $q = \exp(i\pi b^2)$ and real $b^2$.
They also verified that the resulting expressions agree with the fusing
matrix of N=1 Liouville field theory in the Neveu-Schwarz (NS) sector
\cite{Hadasz:2007wi, Chorazkiewicz:2008es}. A central goal of the
present work is to extend the previous expression to include both
NS and Ramond (R) sector fields. The way in which we shall achieve
our goal is quite interesting in its own right.

Let us recall that the expression for the Racah-Wigner symbol found in
\cite{Hadasz:2013bwa} generalized previous formulas by Ponsot and Teschner for
the Racah-Wigner symbol of \Uqtwo\ \cite{Ponsot:1999uf, Ponsot:2000mt}.
In a remarkable recent paper \cite{Teschner:2012em}, Teschner and Vartanov
found an alternative and much more natural way to express the same
Racah-Wigner symbol. In particular, the new formulation is very closely
modeled after the famous expressions for the Racah-Wigner coefficients
of finite dimensional \Uqtwo\ representations \cite{Kirillov:1991ec, Kac:1989ai},
only that an integral appears instead of the usual summation and $q$-factorials are
replaced by double Gamma functions.

Our strategy here is to extend the Teschner-Vartanov expressions for
the Racah-Wigner symbol of \Uqtwo\ to the supersymmetric case. Up
to certain sign factors, this step is relatively straight-forward, taking
into account some of the properties of the formula derived in
\cite{Hadasz:2013bwa}. The resulting expression is so natural that its
extension to the R sector is rather easy to guess. Only the sign factors
are a bit tricky to extend. We shall come up with a concrete proposal.
In order to test our prescription for both NS and R sector labels we
shall continue the integral formulas from spins $\alpha \in Q/2+i
\mathbb{R}$ to the discrete set $j = - \alpha/b \in \mathbb{N}/2$ at
which the integrals can be evaluated by summing over certain residues.

When $j$ is integer, the result of this evaluation gives the known
6J symbols for finite dimensional spin $j$
representations of \Uqosp\ \cite{Minnaert:1994gs, Minnaert:1996gn}.
This limit only uses information from the NS sector, but can be
considered a very strong test of our proposal for the universal
Racah-Wigner symbol, including the sign factors we prescribe in
the NS sector.

In order to probe the R sector of the theory we make use of a remarkable
observation in \cite{Saleur:1989gj, Ennes:1997kx}. These authors found that the
6J symbols for finite dimensional integer spin representations of U$_{q'}$(sl(2))
and \Uqosp\ actually coincide when $q' = i \sqrt{q}$. Because of the
usual relation between the deformation parameter $q= \exp(i\pi/(2k+3)$
and the level $k$, the deformation parameter $q'$ actually tends to
$q'=i $ in the semiclassical limit $k \rightarrow \infty$ of \Uqosp,
i.e.\ it is associated to a point $q' = \exp(i\pi/(k+2)$ with $k=0$,
deeply in the quantum region of U$_{q'}$(sl(2)). In this sense, the
numerical coincidences between 6J symbols of finite dimensional
representations observed in \cite{Saleur:1989gj,Ennes:1997kx} can be
thought of as a non-perturbative duality.~\footnote{We thank Edward
Witten for stressing this aspect of the duality in a private
conversation.} In our context we will find that the limiting \Uqosp\
Racah-Wigner symbols with discrete weights, including those
corresponding to half-integer spin $j$, coincide with the 6J symbols
of finite dimensional representations of U$_{q'}$(sl(2)). Thereby, we
provide  highly non-trivial evidence for our choice of sign factors
in the R sector of the theory.

The tests of our proposal we described in the previous two paragraphs
exhaust the data provided by finite dimensional representations of
deformed universal enveloping algebras. On the other hand, we can
evaluate our proposed Racah-Wigner symbol for a larger set of labels
$\alpha$ which are parametrized  by a pair of spin labels $(j,j')$.
When $j'=0$, we are back to the case discussed above. But for nontrivial
values of $j'$ the limiting value of the Racah-Wigner symbol may be
written as a product of two 6J symbols with different values of $q$.
In reaching such a conclusion, details of the sign factors become even
more crucial. While the result has no direct interpretation in terms of
finite dimensional representation theory of universal enveloping
algebras, it can be understood from the relation between Liouville
theory and minimal models in conformal field theory. Hence it adds
quite significantly to the testing of our main proposal.

The plan of this paper is as follows. In the next section we shall re-address
the case of \Uqtwo\ and show how to recover the Racah-Wigner coefficients
of finite dimensional representations from the formula of Teschner and Vartanov.
After this warm-up, we can turn to the supersymmetric case in section 3. There
we propose a new expression for the Racah-Wigner symbol of \Uqosp. The comparison
with the 6J symbols for integer spin representations of \Uqosp\ and with finite
dimensional representations of \Uqtwo\ is performed in section 4.
We conclude this work with a number of comments on open problems, including some
speculations about the extension of the duality between \Uqtwo\ and \Uqosp\ to
infinite dimensional self-dual representations.

\section{The Racah-Wigner symbol of \Uqtwo}

In this section we will start from a recent integral formula for
the Racah-Wigner symbol of a self-dual series of representations
of \Uqtwo\ with $q=e^{i \pi b^2}$,  parametrized by $\alpha = Q/2
+ i \mathbb{R}, Q = b + b^{-1}$ \cite{Teschner:2012em}.
This symbol turns out to simplify when we consider its analytic
continuation to parameters  $\alpha = -j b - j'b^{-1}; \ j,j'\in
{\mathbb{N} \over 2} $. In fact, it can be then written as a sum
over finitely many pole contributions. We can compare the resulting
expressions  with the formulas for Racah-Wigner coefficients
of finite dimensional representations of \Uqtwo\ and find complete
agreement, at least up to some normalization dependent prefactors.

Let us begin our discussion by reviewing the formulas for the
universal Racah-Wigner coefficients of \Uqtwo\ which were
proposed by Teschner and Vartanov \cite{Teschner:2012em}
\begin{eqnarray}\label{RW_bosonic}
\left\{ \begin{array}{cc  c}
  \alpha_1 & \alpha_3  & \alpha_s \\
 \alpha_2 & \alpha_4  & \alpha_t
  \end{array}  \right\}
  & =& \Delta(\alpha_1, \alpha_2, \alpha_s) \Delta(\alpha_s, \alpha_3, \alpha_4)\Delta(\alpha_t, \alpha_3, \alpha_2)\Delta(\alpha_4, \alpha_t, \alpha_1)
\\ && \nonumber
\times \int_{\mathcal{C}} du \, S_b(u-\alpha_{12s}) \,  S_b(u-\alpha_{s34}) \,
S_b(u-\alpha_{23t}) \,  S_b(u-\alpha_{1t4}) \\
&& \hspace{40pt} \nonumber
S_b(\alpha_{1234} -u ) \,  S_b(\alpha_{st13} -u) \,
S_b(\alpha_{st24} -u) \,  S_b(2Q - u)
\end{eqnarray}
where
\begin{equation}
\label{Delta}
\Delta(\alpha_3, \alpha_2, \alpha_1)  = \left( \frac{ S_b(\alpha_{123} -Q ) }{ S_b(\alpha_{12} - \alpha_3) \, S_b(\alpha_{23} - \alpha_1)  \, S_b(\alpha_{31} - \alpha_2)
} \right)^{\frac12}
\end{equation}
and the multi-index of $\alpha$ denotes summation, e.g. $\alpha_{ij} = \alpha_i +
\alpha_j $. The integral is defined for $\alpha_j = Q/2 + i \mathbb{R}, Q = b +
b^{-1}$ by a contour $\mathcal{C}$ which crosses the real axis in the interval
$(\frac{3Q}{2}, 2Q)$ and approaches $2Q + i \mathbb{R}$ near infinity.
The double sine function $S_b(x)$ is given in terms of  Barnes' double Gamma
function. Its definition and some relevant properties are listed
in appendix \ref{app}. Let us note that Teschner and Vartanov
were able to show that the expression \eqref{RW_bosonic} agrees with
an earlier formula for the Racah-Wigner symbol of \Uqtwo\ that
was established by Teschner and Ponsot \cite{Ponsot:1999uf,Ponsot:2000mt}.
Thus the Racah-Wigner symbol \eqref{RW_bosonic}  coincides with the fusion
matrix of Liouville theory \cite{Teschner:2012em,Teschner:2001rv}. Because
of this relation with conformal field theory (CFT) we shall use some CFT
terminology from time to time. In particular, we  will refer to the labels
$ \alpha_i, \, i=1,\dots,4$ and $\alpha_s, \alpha_t$ as {\em external} and
{\em intermediate} parameters, respectively.

Let us begin our analysis of the Racah-Wigner symbols \eqref{RW_bosonic}
with the prefactor of the integral in the first line. Insertion of the
definition \eqref{Delta} gives
\begin{eqnarray}\label{prefactor}
&& \hspace{-5pt}
\mathcal{P}(\alpha_i) \equiv \Delta(\alpha_1, \alpha_2, \alpha_s) \Delta(\alpha_s, \alpha_3, \alpha_4)\Delta(\alpha_t, \alpha_3, \alpha_2)\Delta(\alpha_4, \alpha_t, \alpha_1)
=
 \\[4mm] \nonumber
&&\hspace{-2pt}
 \left( \frac{ S_b(\alpha_{12s} -Q)  S_b(\alpha_{s34} -Q)
  }{S_b(\alpha_{12} - \alpha_s) \, S_b(\alpha_{2s} - \alpha_1)  \, S_b(\alpha_{1s} - \alpha_2) \,
  S_b(\alpha_{34} - \alpha_s) \, S_b(\alpha_{3s} - \alpha_4)  \, S_b(\alpha_{4s} - \alpha_3)
}
\right)^{\frac12}
\\ \nonumber
&&\hspace{-7pt}
\times
 \left( \frac{ S_b(\alpha_{23t} -Q)  S_b(\alpha_{1t4} -Q)
  }{S_b(\alpha_{23} - \alpha_t) \, S_b(\alpha_{2t} - \alpha_3)  \, S_b(\alpha_{3t} - \alpha_2) \,
  S_b(\alpha_{14} - \alpha_t) \, S_b(\alpha_{1t} - \alpha_4)  \, S_b(\alpha_{4t} - \alpha_1)
}
\right)^{\frac12}\!.
\end{eqnarray}
 We observe that the prefactor vanishes each time one of the external parameters $\alpha_i$
 approaches the so called degenerate value $ \alpha_{n,n'} \equiv - \frac{nb}{2} -
 \frac{ n'}{2b}; \, n,n' \in \mathbb{Z}_{\geq 0}$, and one of the intermediate parameters
  $\alpha_x, (x=s,t)$  satisfies the condition
\begin{eqnarray}\label{fusion_rulesB}
\alpha_x = \alpha_j - \frac{x b }{ 2} - \frac{ x'}{ 2b} \, , \!
 \quad \!
x \in \{ -n, -n + 2, \dots , n \} \, ,
 \quad \!
 x' \in \{ -n', -n' + 2, \dots , n' \}\,
\end{eqnarray}
where the labels $i,j \in \{1,2 \}$ or $ \{3,4 \}$ for $x=s$, and $i,j \in \{2,3 \}$ or $ \{1,4 \}$ for $x=t$.
In Liouville theory, the values  $\alpha_{n,n'}$ are associated with so-called degenerate fields which satisfy
additional null vector decoupling equations. These restrict the possible operator products to a finite set of
terms which are labeled by  parameters satisfying so-called {\em fusion rules}, i.e. conditions of the form
\eqref{fusion_rulesB}.

Let us now consider a limit of the Racah-Wigner symbol where one of the external parameters becomes
degenerate and the intermediate parameter $\alpha_{s}$ satisfies the condition \eqref{fusion_rulesB}.
As we shall show below, the limit is finite and non-zero because the integral
in eq.\ (\ref{RW_bosonic}) contributes singular terms canceling  zeroes from the prefactor.
In order to see how this works in detail, let us focus on the limit
$\alpha_2 \to -\frac{nb }{ 2}$ ($n>0$) and
$ \alpha_s \to \alpha_1 -  \frac{ s b }{ 2}$. The zero in the prefactor
comes from the first two terms in the denominator of eq.\
(\ref{prefactor})
\bea
&& \hspace{-20pt}
\lim_{ \substack{ 
\alpha_2 \to -\frac{nb }{ 2} \\
\alpha_s \to \alpha_1 -  \frac{ s b}{ 2}
}
}
\left(  S_b(\alpha_{12} - \alpha_s) \, S_b(\alpha_{2s} - \alpha_1)  \right)^{-\frac12}
=
\left( S_b \left( \frac{s -n }{ 2} b\right)  S_b\left( - \frac{s +n }{ 2}b \right)  \right)^{-\frac12}
   \\[-4pt]
&& \hspace{80pt}
=
\left( -2 \sin( \pi b^2)  \right)^{\frac{n}{2}} \,
\left( \qb{ \frac{n-s}{2}} \qb{ \frac{n+s}{2}} \right)^\frac12
S_b(0)^{-1}
\eea
where we used the shift relation (\ref{sin_qb}) for the double sine function
and the notation
\begin{equation}\label{x-b}
[x]= \frac{\sin(\pi b^2 x) }{ \sin{\pi b^2} }\ .
\end{equation}
For integer $x$ the factorial $[x]!$ is defined as,
\begin{equation}\label{x-b-factorial}
[x]! = \prod_{a=1}^x [a] = \left( \sin{\pi b^2}\right)^{-x}  \prod_{a=1}^x \sin(\pi b^2 a)\ .
\end{equation}
In order to obtain a finite non-zero limit for the full Racah-Wigner symbol,
the integral must contribute a divergent factor $S_b(0)$ to cancel the
corresponding term from the prefactor. Let us therefore take a closer look
at the integral in eq.\ \eqref{RW_bosonic}.
Its   analytic continuation to $ \alpha_2 = -\frac{nb }{ 2},  \alpha_s = \alpha_1 -
\frac{ s b }{ 2}$ is defined by the same integral with a deformed contour $\mathcal{C}'$,
see figure 1 and figure 2 for the cases $s\geq0$ and $s<0$, respectively. As we deform  the original contour we have to take into account contributions from poles. We shall
split these into two groups and denote them by $I_1, I_2$, respectively,
\begin{eqnarray}\label{integral}
\int_{\mathcal{C}'} du \hspace{-5pt} &&  S_b(u-\alpha_{12s}) \,  S_b(u-\alpha_{s34}) \,
S_b(u-\alpha_{23t}) \,  S_b(u-\alpha_{1t4}) \\
&& \nonumber
S_b(\alpha_{1234} -u ) \,  S_b(\alpha_{st13} -u) \,
S_b(\alpha_{st24} -u) \,  S_b(2Q - u) = I_{\mathrm{reg}} +  I_1 + I_2\ . 
\end{eqnarray}
The first term $I_{\mathrm{reg}}$ denotes the integral over the original contour and a regular contribution. 
The singular terms $I_1$ and $I_2$ will be described and calculated in the next few paragraphs.

By definition, the first singular term $I_1$ has origin in the two double sine functions
$ S_b(u-\alpha_{s34}) \,  S_b(\alpha_{1234} - u )$.
Let us first consider the case of $s \geq 0$. Then the poles of $ S_b(u-\alpha_{s34})$ in
$u = \alpha_{s34}  - p b $ ($ 0 \leq p  \leq \frac{ n-s}{ 2} $)  lie on the left side of
the contour $\mathcal{C}$, see figure 1.  When we deform the contour to $\mathcal{C}'$ we
thus obtain contributions from non-vanishing residues in these points. These residues are
proportional to the other double sine function $ S_b(\alpha_{1234} -  \alpha_{s34}+ p b)$
and in the limit  $ \alpha_2 \to -\frac{nb }{ 2} , \alpha_s \to \alpha_1 - \frac{sb }{ 2}$
become singular. This is  the so called pinching mechanism, see e.g.\ \cite{Ponsot:2000mt},
Lemma 3 and \cite{Hadasz:2004cm,Hadasz:2007wi} for similar calculations. In the end we
obtain the following sum
\begin{eqnarray}\label{sum1}
 \hspace{-5pt}
I_1 & = & \sum_{p=0}^\frac{n-s}{2}
 \!
\Bigg(\!
\frac{ \left( -2 \sin( \pi b^2)  \right)^\frac{s-n}{2}  S_b(0)}{\qb{p} \qb{{n-s \over 2} -p} } \,
 S_b(\alpha_{34} \!-\alpha_{1} +  \frac{n b}{2} \!  - pb) \,
\nonumber
 \\
&&  S_b(\alpha_{1t} \!  - \alpha_4 \! + pb) 
  S_b(\alpha_{14}  \!- \alpha_{t} + \frac{(n-s) b }{ 2} \! - p b )\, S_b(\alpha_{3}\!-\alpha_{t} -\! \frac{s b }{ 2}\! - p b )
 \, \\
\nonumber
&& 
 S_b(\alpha_{t}\! - \alpha_3 - \! \frac{n b }{ 2} + pb)
 S_b(2Q\! - \alpha_{134} + \frac{s b}{ 2} + p b ) \!\! \Bigg).
\end{eqnarray}
\begin{figure}
\centerline{\includegraphics[width=7cm]{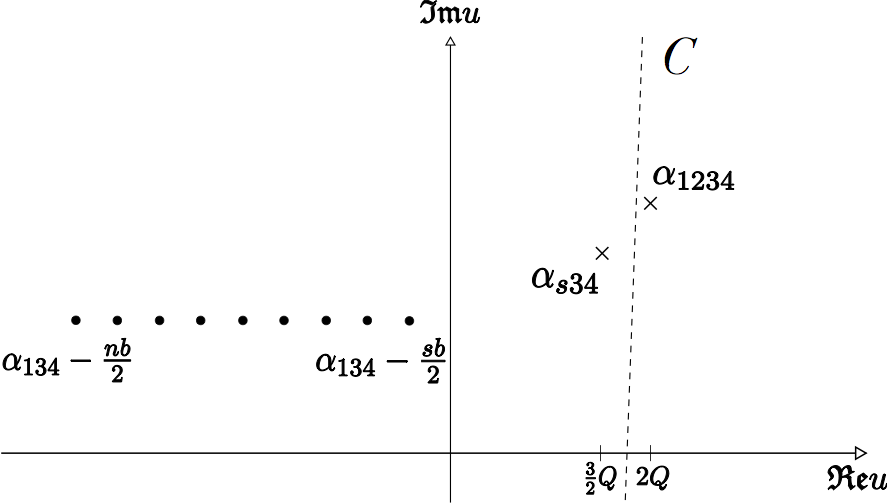}
\quad
\includegraphics[width=7cm]{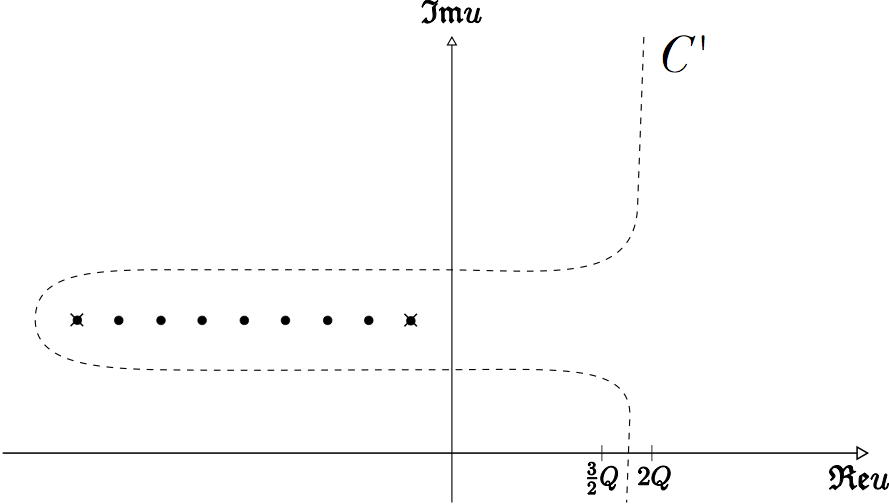}}
\caption{The original integration contour $\mathcal{C}$ passes between the points
$ u = \alpha_{s34}  $ and $u = \alpha_{1234}$. As we deform the contour
to $\mathcal{C}'$,  the  poles contribute to  singular  term $ I_1 $ due to the pinching mechanism.}
\label{fig:1}
\end{figure}

When $s<0$ the function $ S_b(u-\alpha_{s34})$  has poles in  $u = \alpha_{s34}  - p b $ ($- \frac{s}{2}
\leq p \leq \frac{n-s}{ 2}$). In the limit $ \alpha_s \to \alpha_1 -  \frac{ s b }{ 2}$ these are situated
on the left side of the contour $\mathcal{C}$, see figure 2. On the other hand the  function
$ S_b(\alpha_{1234} - u ) $ has poles in $u = \alpha_{1234}  + p b $  ($0\leq p \leq - \frac{s}{2}$)
that are located on the right side of the contour. While deforming  the contour to $\mathcal{C}'$ we
pick up contributions from all these poles. Each residue is proportional to $ S_b(\alpha_{12} -\alpha_{s}
+ pb) $ and develops a singularity in the limit $ \alpha_2 \to -\frac{nb }{ 2} , \alpha_s \to \alpha_1 -
\frac{sb }{ 2}$. The final result will be the same as in the case  (\ref{sum1}) where we 
assumed  $s\geq 0$.
\begin{figure}[b]
\centerline{\includegraphics[width=6cm]{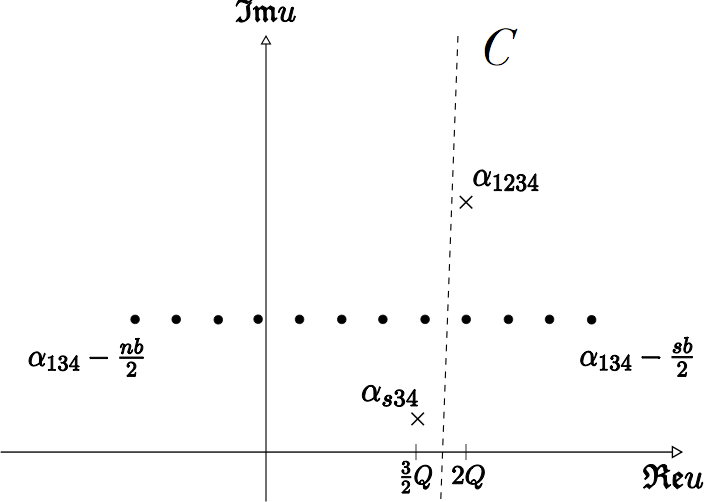}
\includegraphics[width=5.7cm]{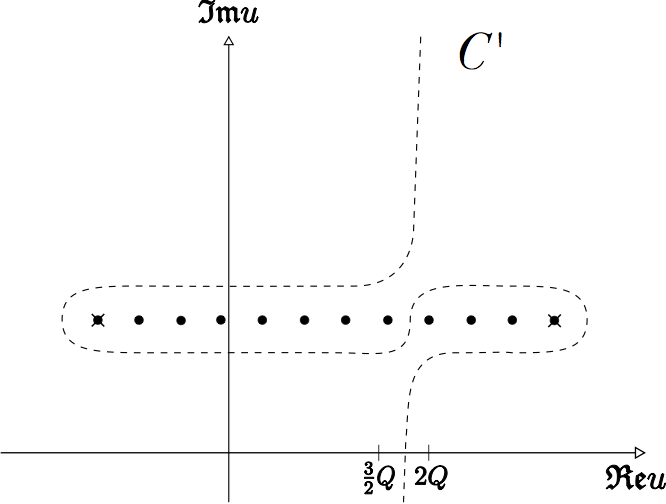}}
\caption{When $s<0$ we have to deform the contour in the above way.
The poles appear on both sides of the
contour $\mathcal{C}$ and  they all  give singular  contribution to $I_1$.}
\label{fig:2}
\end{figure}

The term we have denoted by $I_2$ come from the  poles of the function $  S_b(u-\alpha_{1t4})$
in $u = \alpha_{1t4}-p'b$ for $0 \leq p'  \leq \frac{ n+s}{ 2} $. Since $s > -n$, the poles lie
on the left side of the contour $\mathcal{C}$, independently of the sign of the parameter $s$
(analogous to figure 1). The residues of all poles we pass while deforming the contour are
proportional to $  S_b(\alpha_{st24} - \alpha_{1t4}+p'b) $. In the limit $\alpha_2 \to -
\frac{nb }{ 2} , \alpha_s \to \alpha_1 - \frac{sb }{ 2}$ they contribute to the second sum of
singular terms,
  \begin{eqnarray}\label{sum2} \nonumber
I_2 & = & \sum_{p'=0 }^\frac{n+ s }{ 2} \! \Bigg( \!
\frac{ \left( -2 \sin( \pi b^2)  \right)^{-\frac{n+s}{ 2}}  S_b(0)
}{ \qb{p'} \qb{\frac{n+s }{ 2} -p'} }
S_b(\alpha_{t4} \!-\alpha_{1} \!+ \frac{(s+n) b}{ 2} \!- p'b )\\
& & S_b(\alpha_{14} \!- \alpha_{3}\! + \frac{n b}{ 2} \!  - p' b )  
S_b(\alpha_{t} \!-\alpha_{3} \! +\! {s b \over 2}\! - p'b ) 
S_b(2Q\! - \alpha_{1t4} \!+ p' b )\\
\nonumber
&& 
S_b(\alpha_{3} \!-\alpha_{t} \! -  {n b \over 2} \! + p'b )
 S_b(\alpha_{13} \! - \alpha_4 \! - {s b\over 2} \! + p'b)
  \!\!\Bigg)\! .
\end{eqnarray}
Combining the two divergent terms $I_1, I_2$ given in eqs.\ (\ref{sum1},\ref{sum2})  with 
the prefactor $\mathcal{P}(\alpha_i)$ from eq.\  (\ref{prefactor}) we obtain a finite 
result for the limit,
\begin{eqnarray} \label{PI1I2}
&& \hspace{-10pt}
\lim_{ \substack{ 
\alpha_2 \to -{nb \over 2} \\
\alpha_s \to \alpha_1 -  { s b \over 2}
}
}
\left\{ \begin{array}{cc  c}
  \alpha_1 & \alpha_3  & \alpha_s \\
 \alpha_2 & \alpha_4  & \alpha_t
  \end{array}  \right\}
  =\lim_{  \substack{ 
\alpha_2 \to -{nb \over 2} \\
\alpha_s \to \alpha_1 -  { s b \over 2}
}
}
 \mathcal{P}(\alpha_i) \, (I_1 + I_2)
\\ \nonumber
 &&
\hspace{-10pt}
  = \! \left( \!{
     S_b(\alpha_{14} + \alpha_t -Q) S_b(\alpha_{3} + \alpha_t   - {n b \over 2}  -Q)
 \over
  S_b(\alpha_3 - \alpha_t  - {n b \over 2} )  S_b( \alpha_t - \alpha_3 - {n b \over 2}  )
 S_b(\alpha_{14} - \alpha_t) \, S_b(\alpha_{1t} - \alpha_4)  \, S_b(\alpha_{4t} - \alpha_1)
} \!\right)^{\frac12}
\\ && \nonumber
\hspace{-10pt}
\left(  \!{  \qb{ n-s \over 2} \qb{ n+s \over 2} \,
  S_b(2 \alpha_{1} - { (s+n) b \over 2}  -Q)     S_b(2 \alpha_{134} - { s b \over 2}  -Q)  \over
      S_b(\alpha_{3t} +   {n b \over 2} )
 S_b(2\alpha_{1} + {(n-s) b \over 2} )
S_b(\alpha_{34} \! - \alpha_1 +  { s b \over 2} ) \,S_b(\alpha_{13} \!- \alpha_4 -  { s b \over 2} )
S_b(\alpha_{14}\! - \alpha_3 -  { s b \over 2} )
} \!\right)^{\!\frac12}
\\[6pt]
\nonumber
 && \hspace{-10pt}
 \Bigg\{  \sum_{q=0}^{n-s \over 2}  \!
{ \left( -2 \sin( \pi b^2)  \right)^{s \over 2}
\over \qb{q} \qb{{n-s \over 2} -q} } \,
 S_b(\alpha_{34}-\alpha_{1} +  {n b \over 2}  - qb) \,
S_b(\alpha_{14} - \alpha_{t} + {(n-s) b \over 2} - q b )  \\
\nonumber
&& \hspace{2pt}
S_b(\alpha_{3} \! -\alpha_{t} \!- {s b \over 2} \! - q b )
\, S_b(\alpha_{t} \! - \alpha_3 \! -  {n b \over 2} \! + qb)
 S_b(\alpha_{1t}  \! - \alpha_4 \! + qb)
 S_b(2Q \!- \alpha_{134} \!+ {s b \over 2} \! + \! q b )
\\\nonumber
&& \hspace{-10pt}
+
 \sum_{p'=0 }^{n+ s \over 2} \!
{ \left( -2 \sin( \pi b^2)  \right)^{-{s \over 2}}
\over   \qb{p'} \qb{{n+s \over 2} -p'} } \,
S_b(\alpha_{t4}-\alpha_{1} + {(s+n) b\over 2} - p'b ) \,
S_b(\alpha_{14} - \alpha_{3} + {n b \over 2}   - p' b )
  \\
\nonumber
&& \hspace{-10pt}
S_b(\alpha_{t} \!-\alpha_{3}\!  + {s b \over 2} \!- \! p'b )
S_b(\alpha_{3} \! -\alpha_{t}\!  -  {n b \over 2}\! + \! p'b )
 S_b(\alpha_{13}\!  - \alpha_4 \! - {s b\over 2}\! + \!  p'b)
  S_b(2Q \! - \alpha_{1t4}\! + \! p' b) \!
\Bigg\}  .
\end{eqnarray}
Suppose now that the other intermediate parameter $\alpha_t$ also satisfies condition
(\ref{fusion_rulesB})   i.e. $\alpha_t \to \alpha_3 -  { t b \over 2}$. Then the prefactor
in the formula above gives zero. On the other hand in each term of the sums there are
double poles for
 $ t \in\{  -n +2p, -n+2p+2, \dots ,  s+2p \}$ and
 $  t \in\{  s - 2p', s -2p' +2, \dots ,  n- 2p' \}$
  coming from
$S_b(\alpha_{3}-\alpha_{t} - {s b \over 2} - p b )  \, S_b(\alpha_{t} - \alpha_3  - {n b \over 2}  + pb) $ and  $ S_b(\alpha_{t}-\alpha_{3}  - p'b + {s b \over 2}) \,
S_b(\alpha_{3}-\alpha_{t}  + \alpha_2- p'b ) $, respectively.
The residue for  a given $\alpha_t \to \alpha_3 - {t b \over 2}$ takes the form
\begin{eqnarray*} 
&& \nonumber
\hspace{-10pt}
\underset{\alpha_t \to \alpha_3 - {t b \over 2}}{Res} 
\Bigg( \!  \lim_{ 
\substack{ 
\alpha_2 \to -{nb \over 2} \\
\alpha_s \to \alpha_1 -  { s b \over 2}
}
} \!\!
\left\{ \begin{array}{cc  c}
  \alpha_1 & \alpha_3  & \alpha_s \\
 \alpha_2 & \alpha_4  & \alpha_t
  \end{array}  \right\} \!\! \Bigg)
 \! =\!
\left(\! { S_b(2 \alpha_{1} - { (s+n) b \over 2}  -Q)    S_b(2 \alpha_{3} - { (t+n) b \over 2}  -Q)      \over
 S_b(2\alpha_{1} + {(n-s) b \over 2} ) \,  S_b(2\alpha_{3} + {(n-t) b \over 2} )
} \! \right)^{\! \frac12}
\end{eqnarray*}
\begin{eqnarray}\label{res1}
\nonumber
&&
\sum_{p=max\{0,\frac{t-s}{2}\} }^{min\{{ n-s \over 2}, { n+t \over 2} \} }
\hspace{-3pt}
{ 2
\left( \qb{ n-s \over 2} \qb{ n+s \over 2} \qb{ n-t \over 2} \qb{ n+t \over 2} \right)^\frac12
\over \qb{p} \qb{{n-s \over 2} -p}  \qb{{s-t \over 2} + p}\qb{{n+t \over 2} -p}}
   { S_b(\alpha_{13} \!- \alpha_4 + pb -  { t b \over 2} )
  \over \left( S_b(\alpha_{13} \! - \alpha_4 \! -  { s b \over 2} ) S_b(\alpha_{13} \! - \alpha_4 \! -  { t b \over 2} )  \right)^{\frac12} }
\\[4pt] \nonumber
&&
\hspace{25pt}
  { S_b(\alpha_{34} - \alpha_1 -pb +  { n b \over 2} )
  \over \left( S_b(\alpha_{34} - \alpha_1 +  { s b \over 2} ) \,S_b(\alpha_{34} - \alpha_1 -  { t b \over 2} ) \right)^{\frac12} }
\,
 { S_b(\alpha_{14} - \alpha_3 -pb +  { (n+t-s)  b \over 2} )
  \over \left( S_b(\alpha_{14} \! - \alpha_3 -  { s b \over 2} ) \,S_b(\alpha_{14} \! - \alpha_3 + { t b \over 2} ) \right)^{\frac12} }
  \\[4pt]
&&
\hspace{25pt}
 { \left( S_b( \alpha_{134} - { s b \over 2}  -Q)   S_b( \alpha_{134} - { t b \over 2}  -Q) \right)^{\frac12} \over
 S_b( \alpha_{134} - {s b \over 2} - p b -Q )}
\end{eqnarray}
where we redefined the second summation parameter $p' = p - {t-s \over 2}$  in order to obtain  two identical sums.
Let us denote the residue above as
\begin{eqnarray}\label{j-RW}
\left\{ \begin{array}{cc c}
  \alpha_1  & \alpha_3  & \alpha_1 -  { s b \over 2} \\
 -{nb \over 2} & \alpha_4  & \alpha_3 -  { t b \over 2}
  \end{array}  \right\}'
  \equiv
  \underset{\alpha_t \to \alpha_3 - {t b \over 2}}{Res} \!
\Bigg( \lim_{ \substack{ 
\alpha_2 \to -{nb \over 2} \\
\alpha_s \to \alpha_1 -  { s b \over 2}
}
} 
\left\{ \begin{array}{c cc  }
  \alpha_1 &  \alpha_3 &  \alpha_s \\
 \alpha_2 & \alpha_4  &  \alpha_t
  \end{array}  \right\} \!\! \Bigg) \, .
  \end{eqnarray}
Now one can set  all the other  external parameters $\alpha_i$ ($i=1,3,4$) to degenerate values,
$ \alpha_i \to - j_i b$, $2 j_i \in \mathbb{Z}_{\geq 0}$. In this case,  eq.\ (\ref{res1})
takes the form
\bea
&& \hspace{-20pt}
\left\{ \begin{array}{cccc  c}
  -j_1 b && -j_3 b && -j_1b-{s b\over 2} \\
 -{n b \over 2} && -j_4 b && - j_3 b- {t \over 2} b
  \end{array}  \right\}'
   =
 2  \left(
{ [2j_1 + { s-n \over 2}]! \over  [2j_1 + {n+s \over 2} +1]!}
{ [2j_3 + { t-n \over 2}]! \over  [2j_3 + {n+t \over 2} +1]!}
   \right)^{\frac12}
   \\
   &&
  \sum_{p=max\{ 0, {t-s\over 2} \}}^{min\{ {n-s\over 2} , {t+n \over 2} \} }
     (-1)^{j_1 + j_3 - p+{n +t \over 2} }
   {  \left( \left[ {n-s \over 2}  \right]! \, \left[ {n+s \over 2}  \right]!  \, \left[ {n-t \over 2}  \right]! \, \left[ {n+t \over 2}  \right]! \right)^{\frac12}
   \over
   \left[ p \right]! \, \left[ p+{s-t \over 2}  \right]! \,\left[ {n -s \over 2}-p \right]! \,\left[ {n+t \over 2} -p \right]! \,
   }
      \\&& \hspace{30pt}
   {
   [j_{134}+p + {s \over 2}+1]!
   \over
   \left( \left[ j_{134} +{s \over 2} +1 \right]! \, \left[ j_{134} +{t \over 2} +1 \right]! \right)^{\frac12}
   }
   \,
      {
  \left( \left[ j_{13}-j_4 + {s \over 2}  \right]! \, \left[ j_{13}-j_4 +{t \over 2}  \right]! \right)^{\frac12}
  \over
   \left[ j_{13}-j_4-p + {t \over 2}  \right]!
  }
     \\[4pt]
     && \hspace{30pt}
  {
  \left( \left[ j_{34}-j_1 -{s \over 2}  \right]! \, \left[ j_{34}-j_1 +{t \over 2}  \right]! \right)^{\frac12}
  \over
   \left[ j_{34}-j_1+p -{ n \over 2}  \right]!
  }
  \,
  {
  \left( \left[ j_{14}-j_3 + {s \over 2}  \right]! \, \left[ j_{14}-j_3 -{t \over 2}  \right]! \right)^{\frac12}
  \over
   \left[ j_{14}-j_3+p -{t +n -s\over 2}  \right]!
  }
 \eea
where we assumed that  $\frac{n}{2}- \frac{\alpha_{134}}{b}   = j_{134} + \frac{n}{2} \in \mathbb{N}$ and we
expressed the $S_b$ functions in terms of the $[.]$-factorials \eqref{x-b-factorial}. The minus sign under the
sum comes from the difference in the shift relations (\ref{sin_qb}) concerning $S_b(-xb)$ and $S_b(-xb + Q)$.
Denoting $ j_2 ={n \over 2} , j_s = j_1 +{s \over 2} ,  j_t = j_3+{t\over 2} $ and shifting the summation
parameter to $z = p+  j_{s34} $, one can see our limit coincides with the 6J symbol for finite dimensional
representations of the quantum deformed algebra \Uqtwo,
\begin{eqnarray}
 \left\{ \begin{array}{cc cc c}
  -j_1 b && -j_3 b && -j_s b \\
 -j_2 b && -j_4 b && -j_t b
  \end{array}  \right\}'
    & =&{ (-1)^{j_s+j_t }  ( [2j_s +1]_q [2j_t+1]_q )^{-\frac12}\over 2 \sin( \pi b^2) \sin( -\pi b^{-2})}
     \left( \! \begin{array}{cc  c}
  j_1 & j_2  & j_s \\
 j_3 & j_4  & j_t
  \end{array} \! \right)_{q}
\end{eqnarray}
where the deformation parameter $q$ is given in terms of $b$ as $q=e^{ i \pi b^2}$
and the quantum numbers $[.]_q$ of \Uqtwo\ are equal those defined in eq. \eqref{x-b},
i.e.\
\begin{equation}\label{x-sl2}
[x]_q \equiv {q^{x} - q^{-x} \over q - q^{-1}} = [x]\ .
\end{equation}
Thus we conclude that the residue of the Racah-Wigner coefficient \eqref{j-RW} analytically continued  to
$\alpha_i = -j_i b, \, 2j_i \in \mathbb{Z}_{\geq 0} $ is equivalent to  the  6J symbol of the finite
dimensional representations of the quantum deformed algebra \Uqtwo.

The 6J symbol of finite dimensional representations of U$_q$(sl(2)) is given by the following sum
\cite{Kirillov:1991ec, Kac:1989ai, AlvarezGaume:1988vr}
\begin{eqnarray}\label{6j-sl2}
&& \hspace{-10pt}
\left( \!\! \begin{array}{cc  c}
  j_1 & j_2  & j_s \\
 j_3 & j_4  & j_t
  \end{array} \!\! \right)_q
   =\sqrt{[2j_s+1]_q [ 2j_t +1]_q} \,	 (-1)^{j_{12} - j_{34} - 2j_s}
\\ && \nonumber
 \hspace{-5pt}
\times
\sum_{z\geq 0}  {(-1)^z\,
\Delta_q(j_s, j_2, j_1) \Delta_q(j_s, j_3, j_4)\Delta_q(j_t, j_3, j_2)\Delta_q(j_4, j_t, j_1)
\, [z+1]_q!
 \over [z-j_{12s} ]_q! \, [z-j_{34s} ]_q! \, [ z-j_{14t} ]_q! \, [ z-j_{23t} ]_q!
[ j_{1234} - z]_q! \,  [j_{13st} - z]_q! \, [ j_{24st} - z]_q! }\, .
\end{eqnarray}
Here, the summation extend over those values of $z$ for which all arguments of
the quantum number $[ . ]_q$ are non-negative. In addition we used the shorthand
$$
\Delta_q(a,b,c) = \sqrt{ [-a+b+c]_q! \, [ a-b+c]_q! \, [a+b-c]_q!  / [a+b+c+1]_q! } \, .
$$
It is worth pointing out the similarities between the expressions  \eqref{6j-sl2}
and the original formula \eqref{RW_bosonic}. In passing to eq.\ \eqref{6j-sl2},
the four factors $\Delta$ got replaced by $\Delta_q$ while the eight functions
$S_b$ have contributed the same number of quantum factorials. In addition, the
integration over $u$ became a summation over $z$.

In the above calculation we have restricted $\alpha$ to a subset of degenerate
labels $\alpha = -j b - j'b^{-1}$ with $j'=0$. One may certainly wonder about
the more general case with $j' \neq 0$. It turns out that the corresponding limit
of the Racah-Wigner symbol can still be evaluated using pretty much the same
steps as before. More precisely, we can continue the Racach-Wigner symbol 
\eqref{RW_bosonic} to general degenerate values
\begin{equation}\label{alphaGen}
 \alpha_i \to -j_i b -j'_i b^{-1};  \quad  j,j'\in {\mathbb{Z}_{\geq 0} \over 2}, 
\end{equation}
evaluate the residue at $\alpha_t = \alpha_j-\frac{t}{2} b-\frac{t'}{2} 
b^{-1}$  and restrict the other intermediate parameter $\alpha_s$ to the 
values \eqref{fusion_rulesB}. These steps define the symbol 
\begin{eqnarray}\label{RW_generalB}
&& \hspace{-25pt} \left\{ \hspace{-5pt}  
\begin{array}{cc cc c}
  -j_1 b  -j_1' b^{-1}  && -j_3 b -j_3' b^{-1} && -j_s b -j_s' b^{-1}\\
 -j_2 b -j_2' b^{-1} && -j_4 b -j_4' b^{-1} && -j_t b -j_t' b^{-1}
  \end{array}  \right\}'
 \\ \nonumber
 &&  \equiv
  \lim_{ \substack{ 
\alpha_j \to -j_j b - j'_j b^{-1} \\
\alpha_k \to -j_k b - j'_k b^{-1} \\
\alpha_l \to -j_l b - j'_l b^{-1}
}
} \,
   \underset{\alpha_t \to \alpha_j - \frac{t}{2} b - \frac{t'}{2} b^{-1} }{Res}
\Bigg(  \lim_{ \substack{ 
\alpha_i \to -j_i b - j'_i b^{-1} \\[-2pt]
\alpha_{s} \to \alpha_k - \frac{s}{2} b - \frac{s'}{2} b^{-1}
}
} 
\left\{ \begin{array}{cc  c}
  \alpha_1 & \alpha_3  & \alpha_s \\
 \alpha_2 & \alpha_4  & \alpha_t
  \end{array}  \right\} \Bigg) \, ,
\end{eqnarray}
where
$$
 j_s = j_k + \frac{s}{2}, \quad  j'_s = j'_k + \frac{s'}{2}; \qquad
  j_t = j_j+ \frac{t}{2},  \quad  j'_t = j'_j + \frac{t'}{2}.
$$
Using the properties of double sine functions \eqref{SBxy} and the assumption
$$
j_{1234}, \ j'_{1234} \in \mathbb{Z}_{\geq 0},
$$
one can express the  limit as a product of two 6J symbols of finite dimensional 
representations of the quantum deformed algebra \Uqtwo
\begin{eqnarray} \label{Fusion-6Js}
\nonumber
&& \hspace{-20pt} \left\{ \hspace{-5pt}  
\begin{array}{cc cc c}
  -j_1 b  -j_1' b^{-1}  && -j_3 b -j_3' b^{-1} && -j_s b -j_s' b^{-1}\\
 -j_2 b -j_2' b^{-1} && -j_4 b -j_4' b^{-1} && -j_t b -j_t' b^{-1}
  \end{array}  \right\}'
  \!\!  =\! (-1)^{j_{st} + j'_{st} + 3 j_{1234st} j'_{1234st} - j_{13} j'_{13} - j_{24} j'_{24} - j_{st} j'_{st}   }
     \\&&
     \times
   { ( [2j_s +1]_q [2j_t+1]_q [2j_s' +1]_{q'} [2j_t'+1]_{q'}  )^{-\frac12}\over 2  \sin( \pi b^2) \sin( -\pi b^{-2}) }
     \left(  \begin{array}{cc  c}
  j_1 & j_2  & j_s \\
 j_3 & j_4  & j_t
  \end{array} \! \right)_{q} \
     \left(  \begin{array}{cc  c}
  j_1' & j_2'  & j_s' \\
 j_3' & j_4'  & j_t'
  \end{array} \! \right)_{q'},
\end{eqnarray}
where the deformation parameters assume two different values, namely  
$q=e^{i \pi b^2}$  and $q'=e^{i\pi b^{-2}}$.

As we anticipated in the introduction, the result has an interesting 
CFT interpretation. The limit we consider gives the value of the fusion 
matrix in Liouville theory where all representations are degenerate and 
both intermediate representations satisfy the fusion rules. The resulting
numbers are expected to describe the fusing matrix of Virasoro minimal 
models, at least after continuation of the parameter $b$ to the imaginary 
discrete values $b = i\beta$ with $\beta^2 = \frac{m+1}{m}$. The associated
central charges
$$c = 13 + 6 (b^2 + b^{-2}) \, \to \, 13-6 (\beta^2 + \beta^{-2}),$$
take discrete values with $c< 1$. When parametrized in terms of the integer 
$m$, our parameters $q$ and $q'$ read     
$$
q = e^{-i \pi \beta^2} = e^{-i\pi\frac{m+1}{m}}\quad   , 
\quad  q' =  e^{-i\pi \beta^{-2}} = e^{-i\pi \frac{m}{m+1}}\ .
$$
Since  \Uqtwo \, 6J symbols are invariant with respect to $q \to q^{-1}$, 
we can also use the parameters $q_1 = \exp(i \pi \frac{m+1}{m})$ and  
$q_2 = \exp(i\pi \frac{m}{m+1})$ on the right hand side of eq.\ \eqref{Fusion-6Js}. The result agrees then with 
the fusing matrix of (unitary) minimal models  \cite{Dotsenko:1984ad},  \cite{Furlan:1989ra}, \cite{Felder:1989wv} 
  \footnote{often \Uqtwo \, deformation parameters are defined as   $q = e^{2 i \pi \beta^{ \pm 2}} $ which in our notation is equal to $ q_1^2, q_2^2$}. 
   Thus we have shown that one can recover the 
fusion matrix of minimal models from the Racah-Wigner symbol \eqref{RW_bosonic} .

Given the connection with minimal models, the product structure of our result
\eqref{Fusion-6Js} is easily understood from the famous coset construction,  
$$ {\text{MM}}_k = (\text{SU}(2)_{k} \times \text{SU}(2)_1)
/ \text{SU}(2)_{k+1}\ , 
$$
for Virasoro minimal models. Here the parameter $k$ is related to $m=k+2$ 
by a finite shift. Sectors of the coset theory  are labeled by three integers 
$(2j, 2j',2l)$ where $ 0\leq 2j \leq k$, $ 0\leq 2j' \leq k+1$, $ l = 0, 
\frac12$. The last label does not play a role because it can be set to 
$l=0$ using the so-called field identification symmetry. The two nontrivial 
factors in the fusing matrix are associated with the SU$(2)$ Wess-Zumino-Witten 
(WZW) models at level $k$ and $k+1$. While the SU$(2)_k$ model contributes a 
factor with $\exp(2\pi i/(k+2)) = q_1^2 $, the 6J symbol with $\exp(2\pi i/
(k+3)) = q_2^2$ comes from the SU$(2)$ WZW model at level $k+1$.

\section{The supersymmetric Racah-Wigner symbol}

After our warmup with the Racah-Wigner symbol of the \Uqtwo, we are
now prepared to study its extension to the supersymmetric case. We shall
define the supersymmetric Racah-Wigner symbol in the next few paragraphs
and comment a bit on its relation with N=1 Liouville field theory and
the Racah-Wigner symbol for self-dual representations of \Uqosp. Then
we perform an analysis along the lines of section 2,
i.e.\ we compute the limit of the Racah-Wigner symbol for a discrete set
of representation labels. The interpretation of the results is a bit more
subtle than in the example of \Uqtwo. It has to wait until
section 4.

As a supersymmetric extension of the Racah-Wigner symbol \eqref{RW_bosonic}
we propose the following integral formula
\begin{eqnarray} \label{RW_osp}
 \left\{ \begin{array}{cc  c}
  \alpha_1^{a_1} & \alpha_3^{a_3}  & \alpha_s^{a_s} \\
  \alpha_2^{a_2} & \alpha_4^{a_4}  & \alpha_t^{a_t}
  \end{array}  \right\}^{\nu_3 \nu_4}_{\nu_1 \nu_2}
  \hspace{-6pt} &=& \! \delta_{\sum_i \nu_i =a_s+a_t \, mod \, 2}
   \Delta_{\nu_4}(\alpha_s, \alpha_2, \alpha_1) \Delta_{\nu_3}(\alpha_s, \alpha_3, \alpha_4)\Delta_{\nu_2}(\alpha_t, \alpha_3, \alpha_2)
   \nonumber
   \\[-6pt]
\\[-4pt]   \nonumber
 && \hspace*{-85pt}
\times  \Delta_{\nu_1}(\alpha_4, \alpha_t, \alpha_1)   \int_{\mathcal{C}} du \, \sum_{\nu=0}^1 \Big( (-1)^{X}
 S_{1+\nu+\nu_4+a_s }(u-\alpha_{12s}) \,  S_{1+\nu+\nu_3+a_s }(u-\alpha_{s34})\\
 \nonumber
&& \hspace{-35pt}
S_{1+\nu+\nu_2 + a_t }(u-\alpha_{23t})   \,
 S_{1+\nu+\nu_1+a_t }(u-\alpha_{1t4}) \,
S_{\nu+\nu_1+ \nu_2+a_t }(\alpha_{1234} -u )
\\[2mm]
 \nonumber
&& \hspace{-35pt}
  S_{\nu+\nu_1+ \nu_3+a_2 }(\alpha_{st13} -u) \,
S_{\nu+\nu_1+ \nu_4+a_3 }(\alpha_{st24} -u) \,  S_\nu(2Q - u)
\Big)
\end{eqnarray}
where
\bea
\Delta_\nu (\alpha_3, \alpha_2, \alpha_1)  \!=\!
\left( \!\! { S_{\nu+ \frac12 a_{123}}\!(\alpha_{123} -Q) \over
 S_{\nu+ \frac12(a_{12}-a_3)}\!(\alpha_{12}\! - \! \alpha_3)
  S_{\nu+ \frac12(a_{23}-a_1)}\!(\alpha_{23} \! - \! \alpha_1)
  S_{\nu+ \frac12(a_{31}-a_2)}\!(\alpha_{31} \!-\! \alpha_2)
} \! \right)^{\!\frac12}
\eea
and the contour $\mathcal{C}$, as in the bosonic case, crosses the real axis in the interval $(\frac{3Q}{2}, 2Q)$ and approaches $2Q + i \mathbb{R}$ near infinity. Note
that the arguments $\alpha^a$ of the Racah-Wigner symbol contain a continuous quantum
number $\alpha \in Q/2 + i \mathbb{R}$ along with a superscript $a$ that can take the
values $a=0$ and $a=1$. The discrete label $a$ keeps track on whether the corresponding representation is taken from the Neveu-Schwarz (NS) or Ramond (R) sector, respectively. We will
comment a bit more on this below.
We define  the Racah-Wigner symbol for the discrete labels $a_i$ satisfying the following conditions
\begin{equation}
a_s = a_1+ a_2 = a_3 + a_4 \, mod \, 2, \qquad
a_t = a_1 + a_4 = a_2 + a_3 \, mod \, 2, \qquad
\sum_{i=1}^4 a_i =0 \, mod \, 2,
\end{equation}
otherwise the symbol is set to zero.
The sign factor
\begin{equation} \label{signX}
(-1)^X =
 (-1)^{\nu( a_s \nu_1 + a_1 \nu_3 + a_4 \nu_4 +  a_1 a_s + a_2 a_4 + a_s+a_t )}
\end{equation}
becomes relevant as soon as some of the discrete labels $a_i$ are nonzero. The
supersymmetric  double sine  functions $ S_{\nu}(x)$ with $\nu=0,1$ are
defined in the appendix~\eqref{sSine}.

Before we continue our analysis, let us make a few comments on the status
of the definition \eqref{RW_osp}, its relation with \Uqosp\ and with N=1
Liouville field theory. In recent work, two
of the authors and Leszek Hadasz computed the Racah Wigner symbols for
a certain series of self-dual representations of the quantum enveloping
superalgebra \Uqosp. The arguments of this symbol
assume values $\alpha \in Q/2 + i \mathbb{R}$. Furthermore, the symbol
defined in \cite{Hadasz:2013bwa} was shown to coincide with the fusing
matrix of N=1 Liouville field theory when all field labels are taken
from the NS sector of the model.
The expression in \cite{Hadasz:2013bwa} extends the one found by
Teschner and Ponsot for \Uqtwo. The latter has been rewritten
by Teschner and Vartanov using some highly non-trivial integral
identities. Our symbol \eqref{RW_osp} with $a_i=0$ was defined to
extend the Teschner-Vartanov version of the non-supersymmetric
symbol to \Uqosp. At the moment we cannot prove
that the expression \eqref{RW_osp}, $a_i=0$, agrees with the formula
derived in \cite{Hadasz:2013bwa} simply because we are missing certain
supersymmetric analogues of the integral identities employed in
\cite{Teschner:2012em}. On the other hand our results below make it
seem highly plausible that both formulas agree. In \cite{Hadasz:2013bwa}
no attempt was made to extend the constructions to the R sector of $N=1$
Liouville field theory. It is likely that \Uqosp\
indeed possesses another self-dual series of representations which can
mimic the R sector and that the fusing matrix involving R sector fields
may be obtained from the Racah-Wigner symbol in an extended class of
self-dual representations, but the details have not been worked out.
Here we just make a bold proposal for the extension of the Racah-Wigner
symbol to cases with some $a_i \neq 0$. Our results below strongly
support a relation with the R sector of N=1 Liouville field
theory.

After these comments on the Racah-Wigner symbol \eqref{RW_osp}, we would
like to repeat the analysis we have performed in section 2.
Let us start with the prefactor of our Racah-Wigner symbol. When written in terms of
the double sine function, it takes the from
\begin{eqnarray}\label{prefactor2}
&&
\hspace{-30pt}
\mathcal{P}(\alpha_i, \nu_i) =  \Delta_{\nu_4}(\alpha_s, \alpha_2, \alpha_1) \Delta_{\nu_3}(\alpha_s, \alpha_3, \alpha_4)\Delta_{\nu_2}(\alpha_t, \alpha_3, \alpha_2)\Delta_{\nu_1}(\alpha_4, \alpha_t, \alpha_1)
 \\
 \nonumber
 && \hspace{+6pt}
 =
 \left(  S_{\nu_4+ a_s}(\alpha_{12s} -Q) S_{\nu_3+ a_s}(\alpha_{s34} -Q)
  S_{\nu_2+ a_t}(\alpha_{23t} -Q) S_{\nu_1+ a_t}(\alpha_{14t} -Q)
 \right)^{\frac12}
\\
\nonumber
&&
\hspace{+40pt}
\Big(
 S_{\nu_4}(\alpha_{12} - \alpha_s) \,S_{\nu_4+a_1}(\alpha_{1s} - \alpha_2) \,
 S_{\nu_4+a_2}(\alpha_{2s} - \alpha_1)
 \\ \nonumber
 && \hspace{+45pt}
 S_{\nu_3}(\alpha_{34} - \alpha_s) \, S_{\nu_3+a_4}(\alpha_{s4} - \alpha_3) \,  S_{\nu_3+a_3}(\alpha_{3s} - \alpha_4) \, \\
 \nonumber
 && \hspace{+45pt}
  S_{\nu_2}(\alpha_{23} - \alpha_t) \, S_{\nu_2+a_2}(\alpha_{t2} - \alpha_3) \,  S_{\nu_2+a_3}(\alpha_{3t} - \alpha_2) \,\ \\ \nonumber
 && \hspace{+42pt}\
   S_{\nu_1}(\alpha_{14} - \alpha_t) \,S_{\nu_1+a_1}(\alpha_{1t} - \alpha_4) \,
 S_{\nu_1+a_4}(\alpha_{4t} - \alpha_1)
\Big)^{-\frac12}\ . 
\end{eqnarray}
By analogy with the bosonic case we expect that the prefactor vanishes each 
time one of the external parameters  approaches a degenerate value  $\alpha_{i} = -
\frac{nb}{2} - \frac{n'}{2b}$ and one of the intermediate parameters $\alpha_x, 
(x=s,t)$ satisfies the condition
\begin{eqnarray}\label{fusion_rulesS}
\hspace{-10pt}
\alpha_x = \alpha_j - \frac{x b }{ 2} - \frac{ x'}{ 2b} \, , \!
 \quad \!
x \in \{ -n, -n + 2, \dots , n \} \, ,
 \quad \!
 x' \in \{ -n', -n' + 2, \dots , n' \} ,
\end{eqnarray}
where the labels $i,j \in \{1,2 \}$ or $ \{3,4 \}$ for $x=s$, and $i,j \in \{2,3 \}$ or $ \{1,4 \}$ for $x=t$. Using  properties of supersymmetric double sine functions listed in appendix \ref{app} one can check that the prefactor  indeed has zeroes in these cases, provided that the following conditions are satisfied, 
\begin{eqnarray}\label{nu_s_condition}
\nonumber
&&
 \frac{n-s}{2} + \frac{n'-s'}{2} \in  2 \mathbb{N} +1 +
\left\{ \begin{array}{cccc}
  \nu_4, \quad  &  \mathrm{degenerate}  & \alpha_i, & i= 1,2 \\
 \nu_3, \quad  &  \mathrm{degenerate}  & \alpha_i, & i= 3,4
  \end{array}
\right.
\\[-8pt]
\\
\nonumber
&&
   \frac{n+s}{2} + \frac{n'+s'}{2} \in 2 \mathbb{N} +1 +
\left\{ \begin{array}{cccc}
  \nu_4 + a_i, \quad  &  \mathrm{degenerate}  & \alpha_i, & i= 1,2 \\
 \nu_3+ a_i, \quad  &  \mathrm{degenerate}  & \alpha_i, & i= 3,4\
  \end{array}
\right.
\end{eqnarray}
by the intermediate parameter $\alpha_s$, and
\begin{eqnarray}\label{nu_t_condition}
&& \nonumber
 \frac{n-t}{2} + \frac{n'-t'}{2} \in  2 \mathbb{N} +1 +
\left\{ \begin{array}{cccc}
  \nu_1, \quad  &  \mathrm{degenerate}  & \alpha_i, & i= 1,4 \\
 \nu_2, \quad  &  \mathrm{degenerate}  & \alpha_i, & i= 2,3
  \end{array}
\right.
\\[-8pt]
\\
\nonumber
&&
 \frac{n+t}{2} + \frac{n'+t'}{2} \in 2 \mathbb{N} +1 +
\left\{ \begin{array}{cccc}
  \nu_1 + a_i, \quad  &  \mathrm{degenerate}  & \alpha_i, & i= 1,4 \\
 \nu_2+ a_i, \quad  &  \mathrm{degenerate}  & \alpha_i, & i= 2,3
  \end{array}
\right.
\end{eqnarray}
by  $\alpha_t$.
As one example, let us discuss the condition \eqref{nu_s_condition} and suppose that $\alpha_i =
\alpha_1= -
\frac{nb}{2} - \frac{n'}{2b}$ for definiteness. It follows that $\alpha_j = \alpha_2$ because
$\alpha_1$ and $\alpha_s$ appear only in combination with $\alpha_2$ in
the arguments of the double sine functions. According to eq.\ \eqref{Snu_poles} the first 
double sine function $S_{\nu_4}(\alpha_{12} -\alpha_s)$ runs into a pole provided that 
its argument $\alpha_{12}-\alpha_s = \frac{s-n}{2}b + \frac{s'-n'}{2} b^{-1}$ satisfies $\frac{n-s}{2} +\frac{n'-s'}2 \in 2 \mathbb{N} - 1 + \nu_4$. The second function  $S_{\nu_4+a_1}(\alpha_{1s} - \alpha_2)$  has a pole if $\frac{n+s}{2} +\frac{n'+s'}
2 \in 2 \mathbb{N} - 1 + \nu_4 + a_1$. If both conditions are fulfilled the 
prefactor become zero. Let us note  that  this can be the case only if
$ s+s' \in 2 \mathbb{Z}_{\geq0}+ a_1$ and equivalently, due to eq.\ \eqref{fusion_rulesS},  
$ n+n' \in 2 \mathbb{Z}_{\geq0}+ a_1$. The analysis for the other cases is similar.

In general, the conditions \eqref{nu_s_condition}, \eqref{nu_t_condition} can be 
satisfied only if degenerate parameters are of the form
\begin{equation}\label{alpha-a}
\alpha_i = -\frac{nb}{2} - \frac{n'}{2b} , \qquad n+ n' \in 2 \mathbb{Z}_{\geq0}+ a_i.
\end{equation}
This reflects the situation in the  $N=1$ Liouville field theory, where degenerate 
representations in the NS and R sectors  are labeled by $\alpha_{n,n'}$ with  even  
and  odd  $n+n'$, respectively. Additionally, the pattern of zeroes of the prefactor 
$ \mathcal{P}(\alpha_i, \nu_i)$  well matches with  fusion rules of $N=1$ Liouville 
field theory. This provides a first non-trivial test for our proposal.

We plan to test our proposal (\ref{RW_osp}) further by continuing it to degenerate
parameters, as in the previous section. To this end, let us consider the limit  of the
Racah-Wigner symbol
where $\alpha_2 \to -{n b\over 2} $, $ \alpha_s \to \alpha_1 - {s \over 2}$ and the conditions \eqref{fusion_rulesS}- \eqref{alpha-a} are satisfied. Before talking the limit it is useful to pass from the summation over $\nu$ to
a new summation index $\nu' = \nu + \nu_3 + a_s$. The Racah-Wigner symbol
then reads,
\begin{eqnarray}\label{RW_s_a2}
\nonumber
 \left\{ \begin{array}{cc  c}
  \alpha_1^{a_1} & \alpha_3^{a_3}  & \alpha_s^{a_s} \\
  \alpha_2^{a_2} & \alpha_4^{a_4}  & \alpha_t^{a_t}
  \end{array}  \right\}^{\nu_3 \nu_4}_{\nu_1 \nu_2}
 \hspace{-10pt} & =& \hspace{-1pt} \delta_{\sum_i \nu_i =a_s+a_t  \, mod \, 2}
\, \mathcal{P}(\alpha_i, \nu_i)
\int_{\mathcal{C}} du \,
\sum_{\nu'=0}^1
\Big(
  (-1)^{X} \,  S_{1+\nu_3 + \nu_4 + \nu' }(u-\alpha_{12s})
\\[1mm]
&&  \hspace{-80pt}
  S_{1+ \nu' }(u-\alpha_{s34})
S_{1+\nu_1+\nu_4+ \nu' }(u-\alpha_{23t})
 S_{1+\nu_2+\nu_4+ \nu' }(u-\alpha_{1t4})  \,
S_{\nu_4+ \nu' }(\alpha_{1234} -u ) \\[2mm]
  \nonumber
&&  \hspace{-80pt}
 S_{\nu_1 + \nu' + a_1}(\alpha_{st13} -u)
S_{\nu_2 + \nu' + a_2 }(\alpha_{st24} -u) \,  S_{\nu_3 + \nu' + a_s}(2Q - u) \Big)\ .
\end{eqnarray}
As in the previous section, we need to determine the singular contributions
from the integral
\bea
&&  \hspace{-20pt}
\int_{\mathcal{C}'} du \,
\sum_{\nu'=0}^1
\Big(
  (-1)^{X} \,  S_{1+\nu_3 + \nu_4 + \nu' }(u-\alpha_{12s})
    S_{1+ \nu' }(u-\alpha_{s34})
S_{1+\nu_1+\nu_4+ \nu' }(u-\alpha_{23t})
\\[1mm]
&&  \hspace{10pt}
 S_{1+\nu_2+\nu_4+ \nu' }(u-\alpha_{1t4}) \,
S_{\nu_4+ \nu' }(\alpha_{1234} -u )
 S_{\nu_1 + \nu' + a_1}(\alpha_{st13} -u)
 S_{\nu_2 + \nu' + a_2 }(\alpha_{st24} -u)  \\[2mm]
  \nonumber
&&  \hspace{10pt}
\,  S_{\nu_3 + \nu' + a_s}(2Q - u) \Big)
=
I'_{\mathrm reg} + I'_1 + I'_2\ . 
\eea
Note that the product
$  S_{1+\nu' }(u-\alpha_{s34}) S_{\nu_4 + \nu'  }(\alpha_{1234} -u )  $
has poles in the positions
$u = \alpha_{134} - {s b \over 2} - p b $ for $ p \in \{\nu',\nu'+ 2, \dots,   { n-s \over 2} - \nu'\} $ ($\nu'$ keeps track of the parity of $p$).
 Due to  the ``pinching mechanism'' each pole contributes a  singular term. Once we include the summation over $\nu'=0,1$, the sum of singular terms runs through all values of $p \in \{0,1, \dots, {n-s \over 2} \}$,
 \begin{eqnarray*}
 &&  \hspace{-15pt} 
 I'_1 = \sum_{p=0}^{ n-s \over 2 }
  (-1)^{X}
 {  \left( 2 \cos(\frac{\pi b^2}{2} )\right)^{\frac{s-n}{2}} S_{1}(0)
\over  \qn{p}! \qn{{n-s \over 2} -p }!  } \,
S_{1+\nu_3 + \nu_4 + \nu' }(\alpha_{34}-\alpha_{1} + {nb\over 2} - pb)
\\ 
&&  \hspace{-15pt}
S_{1+\nu_1+\nu_4+ \nu' }(\alpha_{14} - \alpha_{t} + {(n-s) b \over 2} -  pb )
 S_{1+\nu_2+\nu_4+ \nu' }(\alpha_{3}-\alpha_{t} - {s b \over 2} - pb )
 \\
&&  \hspace{-15pt}
 S_{\nu_1 + \nu' + a_1}(\alpha_{1t}   - \alpha_4 + pb)
S_{\nu_2 + \nu' +a_2}(\alpha_{t} - \alpha_3 -{nb \over 2} + pb)
  S_{\nu_3 + \nu' + a_s}(2Q - \alpha_{134} + {s b \over 2} + p b ) ,
 \end{eqnarray*}
 where we used the shift relations for the supersymmetric double sine function (\ref{Sx}) and the notation
 \begin{eqnarray}\label{qsN}
\qn{n}! =
\left\{
\begin{array}{l}
\prod_{j =1 \, mod \, 2}^{n-1}  \cos(j { \pi b^2 \over 2})
\prod_{j =2 \, mod \, 2}^{n}
 \sin( -j {\pi b^2 \over 2})
 \, \left(  \cos({\pi b^2 \over 2}) \right)^{-n} , \ \mathrm{for} \, n\in 2\mathbb{N}
\\[8pt]
 \prod_{j =1 \, mod \, 2}^{n}  \cos( j{ \pi b^2 \over 2} )
\prod_{j =2 \, mod \, 2}^{n-1}
 \sin( -j {  \pi b^2 \over 2})
 \, \left(  \cos({\pi b^2 \over 2}) \right)^{-n} , \ \mathrm{for} \, n\in 2\mathbb{N} +1 \ .
\end{array}
\right.
\end{eqnarray}
  With the help of conditions (\ref{nu_s_condition})  one can verify that the functions
  $ S_{1+\nu_2+\nu_4+ \nu' }(u-\alpha_{1t4}) \, S_{\nu_2 + \nu' + a_2 }(\alpha_{st24} -u) $
  have  poles located in $ u = \alpha_{1t4}-p'b   $, where  $ p' \in \{\mu, \mu+ 2, \dots,   { n+s \over 2} - \mu \}, \mu = \nu_2+\nu_4+ \nu'$ mod $2$. They lead to the second sum
  of singular terms $I_2$,
  \begin{eqnarray*}
 &&  \hspace{-20pt} 
 I'_2 =
 \sum_{p'=0}^{ n+s \over 2 }
 {   (-1)^{X} \left( 2 \cos(\frac{\pi b^2}{2} )\right)^{-\frac{n+s}{2}} S_{1}(0)
\over  \qn{p'}! \qn{{n+s \over 2} -p' }!  }
S_{\nu_4 + \nu' }(\alpha_{3} \!- \alpha_t \!-  {nb \over 2} \! + p'b)
 \\
&&  \hspace{-20pt}
S_{1+\nu_1+\nu_4+ \nu' }(\alpha_{14} \! - \alpha_{3} \! + {n b \over 2} \! - p' b )
S_{1+ \nu' }(\alpha_{t}\! -\! \alpha_{3}\!  + {s b \over 2}\!  - p' b )
 S_{\nu_1 + \nu' + a_1}(\alpha_{13} \!-\! \alpha_4 \! + p'b\!  - {s b \over2} )
\\
&&  \hspace{-20pt}
 S_{1+\nu_3 + \nu_4 + \nu' }(\alpha_{t4}\!-\! \alpha_{1} \! +{(n+s)b \over 2} \! - p'b)
  S_{\nu_3 + \nu' + a_s}( 2Q - \alpha_{1t4}  + p' b )\ .
 \end{eqnarray*}
   Once the two singular contributions from the integral are  multiplied by the vanishing prefactor, they give a finite result for the limit
  of the Racah-Wigner symbol,
\begin{eqnarray}\label{res_in_at}
&&  \hspace{-30pt}
\lim_{ \substack{ 
\alpha_2 \to -{nb \over 2} \\
\alpha_s \to \alpha_1 -  { s b \over 2}
}
}
 \hspace{-1pt}
\left\{ \begin{array}{cc  c}
  \alpha_1^{a_1} & \alpha_3^{a_3}  & \alpha_s^{a_s} \\
 \alpha_2^{a_2} & \alpha_4^{a_4}  & \alpha_t^{a_t}
  \end{array}  \right\}^{\nu_3 \nu_4}_{\nu_1 \nu_2}
   \hspace{-2pt}
  =
  \delta_{\sum_i \nu_i =a_s+a_t \, mod \, 2}
\hspace{-5pt}   \lim_{ \substack{ 
\alpha_2 \to -{nb \over 2} \\
\alpha_s \to \alpha_1 -  { s b \over 2}
}
}
 \mathcal{P}(\alpha_i,\nu_i) \, (I'_1 + I'_2)\ . 
\end{eqnarray}
The limit above, similar as in the bosonic case \eqref{PI1I2}, has simple poles when the second intermediate parameter  $\alpha_t \to \alpha_3 - {t b \over 2}$ satisfies the conditions \eqref{fusion_rulesS}, \eqref{nu_t_condition}.
The residue is  given by the following formula,
\begin{eqnarray}\label{int_general}
&&  \hspace{-20pt}
\underset{\alpha_t \to \alpha_3 - {t b \over 2}}{Res}
\Bigg(  \lim_{ \substack{ 
\alpha_2 \to -{nb \over 2} \\
\alpha_s \to \alpha_1 -  { s b \over 2}
}
}
\left\{ \begin{array}{cc  c}
  \alpha_1^{a_1} & \alpha_3^{a_3}  & \alpha_s^{a_s} \\
 \alpha_2^{a_2} & \alpha_4^{a_4}  & \alpha_t^{a_t}
  \end{array}  \right\}^{\nu_3 \nu_4}_{\nu_1 \nu_2}
 \Bigg)
= \delta_{\sum_i \nu_i =a_s+a_t \, mod \, 2}
\\  \nonumber
&&
\hspace{-15pt}
2\,  \left( {  S_{\nu_4+a_s}(2 \alpha_{1} - { (s+n) b \over 2}  -Q)    S_{\nu_2+a_t}(2 \alpha_{3} - { (t+n) b \over 2}  -Q)      \over
 S_{\nu_4+a_1}(2\alpha_{1} + {(n-s) b \over 2} ) \,  S_{\nu_2+a_3}(2\alpha_{3} + {(n-t) b \over 2} )
} \right)^{\frac12}
\\  \nonumber
&&
\hspace{-15pt}
\sum_{p=max\{0,\frac{t-s}{2}\} }^{min\{{ n-s \over 2}, { n+t \over 2} \} }
\Bigg\{
(-1)^{X} \,
   { \left( S_{\nu_3+a_s}( \alpha_{134} - { s b \over 2}  -Q)   S_{\nu_1+a_t}( \alpha_{134} - { t b \over 2}  -Q) \right)^{\frac12} \over
 S_{\nu_3+\nu'+a_s}( \alpha_{134} - {s b \over 2} - p b -Q )}
\\[4pt] \nonumber
&& \hspace{-15pt}
\frac{\left( \qn{n-s \over  2}! \qn{n+s \over 2}!
\qn{n-t \over  2}! \qn{n+t \over 2}! \right)^{\frac12}
\,}{
\qn{p}! \qn{{n-s \over 2} -p }! \qn{p+ {s-t \over 2}}! \qn{{t+n \over 2} -p }!  }
  { S_{\nu_1+\nu'+a_1}(\alpha_{13} - \alpha_4 + pb -  { t b \over 2} )
  \over \left( S_{\nu_3+a_3}(\alpha_{13} \! - \alpha_4 \! -  { s b \over 2} )
  S_{\nu_1+a_1}(\alpha_{13} \! - \alpha_4 \! -  { t b \over 2} )  \right)^{\frac12} }
\\[4pt] \nonumber
&& \hspace{-15pt}
  { S_{1+\nu_3+\nu_4+\nu'}(\alpha_{34} \! - \alpha_1  -pb +  { n b \over 2} )
  \over \left( S_{\nu_3}(\alpha_{34} \! - \! \alpha_1\!  +  { s b \over 2} )
  S_{\nu_1+a_4}(\alpha_{34} \! - \! \alpha_1\!  -  { t b \over 2} ) \right)^{\frac12} }
 { S_{1+\nu_1+\nu_4+\nu'}(\alpha_{14} \! - \alpha_3 -pb +  { (n+t-s)  b \over 2} )
  \over \left( S_{\nu_3+a_4}(\alpha_{14} \! - \! \alpha_3 \!  -  { s b \over 2} )
  S_{\nu_1}(\alpha_{14}\!  - \! \alpha_3\!  + { t b \over 2} ) \right)^{\frac12} } \!
 \Bigg\}.
\end{eqnarray}
In complete analogy to the bosonic case, see eq.\ \eqref{j-RW}, we shall
denote the residue by
\begin{eqnarray}\label{resS}
 \hspace{-10pt}
\left\{  \hspace{-6pt} \
 \begin{array}{cc  c}
  \alpha_1^{a_1} & \alpha_3^{a_3}  & \left(\alpha_1- {s b \over 2}\right)^{a_s} \\
 -{ n b \over 2} & \alpha_4^{a_4}  & \left(\alpha_3- {t b \over 2}\right)^{a_t}
  \end{array}   \right\}^{\!\! \prime \ \nu_3 \nu_4}_{\nu_1 \nu_2}
  \hspace{-5pt}
    \equiv
   \underset{\alpha_t \to \alpha_3 - {t b \over 2}}{Res}
\Bigg(  \lim_{\substack{ 
\alpha_2 \to -{nb \over 2} \\
\alpha_s \to \alpha_1 -  { s b \over 2}
}
}
\left\{ \begin{array}{cc  c}
  \alpha_1^{a_1} & \alpha_3^{a_3}  & \alpha_s^{a_s} \\
 \alpha_2^{a_2} & \alpha_4^{a_4}  & \alpha_t^{a_t}
  \end{array}  \right\}^{\nu_3 \nu_4}_{\nu_1 \nu_2}
 \Bigg) ,
  \end{eqnarray}
where we assume $n \in 2\mathbb{Z}_{\geq0} + a_2$, according to the condition \eqref{alpha-a}. Now we can send all the other external parameters to degenerate values,
$$
\alpha_i \to - j_i b, \qquad 2 j_i \in 2 \mathbb{Z}_{\geq0}+ {a_i }.
$$
Using  the shift relations (\ref{Sx}) for double sine functions  one obtains
\begin{eqnarray}\label{RW_res}
&&  \hspace{-7pt}
 \left\{ \begin{array}{cc cc c}
  -j_1 b && -j_3  b && -j_s b \\
 -j_2b && -j_4 b  && -j_t b
  \end{array}  \right\}^{\!\! \prime \ \nu_3 \nu_4}_{\nu_1 \nu_2}
  \hspace{-3pt}
= \delta_{\sum_i \nu_i =2(j_s+j_t) \, mod \, 2}
\, {(-1)^{A(j_i)} \over  2\cos{ ({\pi b^2\over 2} ) } \cos{ ({\pi  \over 2 b^2} ) }  }
   \\[4pt] \nonumber
   && \hspace{-2pt}
\sum_{z\geq 0} \!
{
(-1)^{X}  (-1)^{ \frac12 z (z-1)  } \qn{z+1}!
\, \Delta_b(j_s, j_2, j_1) \Delta_b(j_s, j_3, j_4)\Delta_b(j_t, j_3, j_2)\Delta_b(j_4, j_t, j_1)
\over
\qn{z-j_{12s}}! \, \qn{z-j_{34s} }! \, \qn{ z-j_{14t} }!
 \qn{z-j_{23t}}!
\qn{ j_{1234} - z}! \,  \qn{j_{13st} - z}! \, \qn{ j_{24st} - z}!
}
\end{eqnarray}
where we denoted
$
{n \over 2} = j_2, \  {s \over 2}= j_s- j_1, \
 {t \over 2} =j_t - j_3
$
and besides  conditions  (\ref{nu_s_condition}), \eqref{nu_t_condition} we assume additionally
\begin{equation}\label{j1234}
 j_{1234}  \in 2 \mathbb{N} + \nu_3 + \nu_4 + a_s ,  \quad \mathrm{and} \quad
 j_{1234} \in 2 \mathbb{N} + \nu_1 + \nu_2 + a_t.
\end{equation}
The sum in \eqref{RW_res} runs over $z= p+ j_{s34}$ such that all arguments $\qn{.}$ are non-negative, and
$$
\Delta_b(a,b,c) = \sqrt{ \qn{-a+b+c}! \, \qn{a-b+c}! \, \qn{a+b-c}!  / \qn{a+b+c+1}! }\, .
$$
The sign $(-1)^{A(j_i)} $ in the prefactor comes from the identity (\ref{Sx}) applied to the terms $S_\nu(-xb - Q)$,
\bea
(-1)^{A(j_i)} &=& (-1)^{\frac14 j_{12s}(j_{12s} -1) + \frac14 j_{s34}(j_{s34} -1) +
 \frac14 j_{23t}(j_{23t} -1) + \frac14 j_{14t}(j_{14t} -1)  +1} \ .
\eea
This concludes our computation of the Racah-Wigner symbol \eqref{RW_osp}
for degenerate labels
$
\alpha_i \to - j_i b, \, 2 j_i \in 2 \mathbb{Z}_{\geq0}+ {a_i }.
$

Let us finally mention that along the same lines one can calculate more general limit of the  Racah-Wigner symbol where the parameters take degenerate values,
 \begin{equation}\label{jj'_S}
\alpha_i \to - j_i b - j_i' b^{-1}, \qquad j_i + j_i' \in \mathbb{Z}_{\geq0} + {a_i \over 2}
 \end{equation}
and the relations (\ref{nu_s_condition}), \eqref{nu_t_condition}  and
\begin{equation}
 j_{1234} + j'_{1234} \in 2 \mathbb{Z}_{\geq0}+ \nu_3 + \nu_4 + a_s ,  \quad \mathrm{and} \quad
 j_{1234} + j'_{1234}\in 2 \mathbb{Z}_{\geq0} + \nu_1 + \nu_2 + a_t
\end{equation}
are assumed. The limit is defined analogously to eqs.\ \eqref{resS} and \eqref{RW_generalB},
\begin{eqnarray*}
&& \hspace{-15pt} \left\{ \hspace{-5pt}  
\begin{array}{cc cc c}
  -j_1 b  -j_1' b^{-1}  && -j_3 b -j_3' b^{-1} && -j_s b -j_s' b^{-1}\\
 -j_2 b -j_2' b^{-1} && -j_4 b -j_4' b^{-1} && -j_t b -j_t' b^{-1}
  \end{array}  \right\}^{\!\! \prime \ \nu_3 \nu_4}_{\nu_1 \nu_2}
 \\ \nonumber
 &&  \equiv
  \lim_{\substack{ 
\alpha_j \to -j_j b - j'_j b^{-1} \\
\alpha_k \to -j_k b - j'_k b^{-1} \\
\alpha_l \to -j_l b - j'_l b^{-1} 
}}\,
   \underset{\alpha_t \to \alpha_j - \frac{t}{2} b - \frac{t'}{2} b^{-1} }{Res}
\Bigg( \! \lim_{ \substack{ 
\alpha_i \to -j_i b - j'_i b^{-1} \\[-2pt]
\alpha_{s} \to \alpha_k - \frac{s}{2} b - \frac{s'}{2} b^{-1}
}} \!
\left\{ \begin{array}{cc  c}
  \alpha_1^{a_1} & \alpha_3^{a_3}  & \alpha_s^{a_s} \\
 \alpha_2^{a_2} & \alpha_4^{a_4}  & \alpha_t^{a_t}
  \end{array}  \right\}^{\nu_3 \nu_4}_{\nu_1 \nu_2}  \Bigg) \, ,
\end{eqnarray*}
where
$$
 j_s = j_k + \frac{s}{2}, \quad  j'_s = j'_k + \frac{s'}{2}; \qquad
  j_t = j_j+ \frac{t}{2},  \quad  j'_t = j'_j + \frac{t'}{2}.
$$
Using  the identity \eqref{SnuXY} for double sine functions  
$S_\nu(-x b -yb^{-1} )$ one may obtain
\begin{eqnarray} \label{RW_Sgen}
&&  \hspace{-20pt} \nonumber 
 \left\{ \begin{array}{cc cc c}
  -j_1 b -j'_1 b^{-1} && -j_3  b  -j'_3 b^{-1}  && -j_s b  -j'_s b^{-1}\\
 -j_2b  -j'_2 b^{-1} && -j_4 b  -j'_4 b^{-1}  && -j_t b  -j'_t b^{-1}
  \end{array}  \right\}^{\!\! \prime \ \nu_3 \nu_4}_{\nu_1 \nu_2}
  \hspace{-3pt}
\sim \delta_{\sum_i \nu_i =2(j_s+j_t) \, mod \, 2}
\\ \nonumber
&& \hspace{-10pt}
\sum_{z\geq 0} \sum_{z'\geq 0}  (-1)^{X}
(-1)^{ \frac12 z (z-1)+ \frac12 z' (z'-1)} (-1)^{B}
 [z+1]_b! \qbm{z'+1}!
 \\ && 
  \Big([z-j_{12s} ]_b! \, [z-j_{34s} ]_b!  [ z-j_{14t} ]_b! \,
 [ z-j_{23t} ]_b!
[ j_{1234} - z]_b! \,   \Big)^{-1}
\\ && \nonumber
 \Big([j_{13st} - z]_b! \, [ j_{24st} - z]_b! \qbm{z'-j'_{12s}}! \, \qbm{z'-j'_{34s} }!  \qbm{ z'-j'_{14t} }! \Big)^{-1} \\ && \nonumber 
 \Big( \qbm{ z'-j'_{23t} }!
\qbm{ j'_{1234} - z'}! \,  \qbm{j'_{13st} - z'}! \, \qbm{ j'_{24st} - z'}! \Big)^{-1}\ . 
\end{eqnarray}
The result is similar to  eq.\ \eqref{RW_res}, with the difference that now we have two sets of brackets $\qn{x}, \qbm{y}$  defined by the formula \eqref{qsN} and the analogous one with $b$ exchanged for $b^{-1}$. Moreover an additional sign comes from eq.\ \eqref{SnuXY},
\begin{equation}\label{-1B}
(-1)^B = (-1)^{ - 2 z j'_{1234st} - 2 z' j_{1234st} } (-1)^{ \frac12 \sum_{i=1}^7 \mu_i (z- x_i)^2 - \nu ( \frac{z^2}{2} -z)}
\end{equation}
 where
 \bea
 \mu_1 = 1+ \nu + \nu_4 + a_s \ \mathrm{mod} \ 2, \   x_1 = j_{12s};
 &\quad&
  \mu_5 =  \nu +\nu_1+ \nu_2 + a_t \ \mathrm{mod} \ 2, \   x_5 = j_{1234};
 \\
 \mu_2 = 1+ \nu + \nu_3 + a_s \ \mathrm{mod} \ 2,  \  x_2 = j_{s34};
  &\quad&
 \mu_6 = \nu +\nu_1+ \nu_3 + a_2 \ \mathrm{mod} \ 2,   \ x_6 = j_{st13};
 \\
  \mu_3 = 1+ \nu + \nu_2 + a_t \ \mathrm{mod} \ 2, \  x_3 = j_{23t};
 &\quad&
   \mu_7 = \nu +\nu_1+ \nu_4 + a_3  \ \mathrm{mod} \ 2,  \  x_7 = j_{st24};
 \\
 \mu_4 = 1+ \nu + \nu_1 + a_t \ \mathrm{mod} \ 2,   \ x_4 = j_{1t4};
 &\quad&
\eea
The final formulas \eqref{RW_res}, \eqref{RW_Sgen} look somewhat similar to the
corresponding equations in section 2. We are now going to see that they are indeed very closely related.

\section{Comparison with the finite dimensional 6J symbols}

Our formulas \eqref{RW_res},  \eqref{RW_Sgen} for the limiting value of the proposed
Racah-Wigner symbol could turn into a strong test of eq.\ \eqref{RW_osp}
provided we were able to show that the expressions \eqref{RW_res},  \eqref{RW_Sgen} give
rise to a solution of the pentagon equation. In our discussion of the
Racah-Wigner symbol for \Uqtwo\ this followed from the comparison
with the 6J symbols for finite dimensional representations. By
construction, the latter are known to satisfy the pentagon equation.
By analogy one might now hope that the coefficients \eqref{RW_res}, \eqref{RW_Sgen}
coincide with the 6J symbols for finite dimensional representations
of the quantum universal enveloping algebra \Uqosp. This,
however, is not quite the case. To start the comparison, we quote
an expression for the 6J symbols of \Uqosp\ from
\cite{Minnaert:1994gs, Minnaert:1996gn},
\begin{eqnarray}
 && \hspace{-20pt} \label{6j-osp}
\left[ \begin{array}{cc  c}
  l_1 & l_2  & l_s \\
l_3 & l_4  & l_t
  \end{array}  \right]_q
   =
    (-1)^{\frac12 (l_{1234} + l_s+l_t)( l_{1234} + l_s+l_t +1) + \frac12 \left( \sum_{i=1}^4 l_i(l_i -1) + l_s(l_s-1) + l_t(l_t-1) \right)  }
\\ &&
\nonumber
\sum_{z\geq 0}
{ (-1)^{ \frac12 z (z-1)  } [z+1]_q'!
  \Delta'_q(l_s, l_2, l_1) \Delta'_q(l_s,l_3, l_4)\Delta'_q(l_t, l_3, l_2)\Delta'_q(l_4, l_t, l_1)
\over
[z-l_{12s} ]_q'! \, [z-l_{34s} ]_q'! \, [ z-l_{14t} ]_q'!
 [ z-l_{23t} ]_q'!
[ l_{1234} - z]_q'! \,  [l_{13st} - z]_q'! \, [ l_{24st} - z]_q'! }
\end{eqnarray}
where  the sum extend over those values of $z$ for which all arguments of
the quantum number $[ . ]_q'$ are non-negative and
$$
\Delta'_q(a,b,c) = \sqrt{ [-a+b+c]_q'! \, [ a-b+c]_q'! \, [a+b-c]_q'!  / [a+b+c+1]_q'! } \, .
$$
Let us stress that irreducible finite dimensional representations of
\Uqosp\ are labeled by integers $l$. Hence all the arguments
$l_i$ in the above 6J symbols satisfy $l_i \in \mathbb{Z}_\geq0$. In the
previous definition the q-number $[.]'_q$ is defined as
\begin{equation}\label{qn_osp}
 [ n ]_q' = { q^{-{n \over 2}} -  (-1)^n q^{n \over 2}  \over
 q^{-{1\over 2}} +  q^{1 \over 2}  }\ .
\end{equation}
For $ q = e^{ i \pi b^2}$ the quantum factorial takes the form
\begin{eqnarray*}
[ n ]_q'! \!=\!
\left\{
\begin{array}{l}
\prod_{j =1 \, mod \, 2}^{n-1}  \cos(j { \pi b^2 \over 2})
\prod_{j =2 \, mod \, 2}^{n}
 \left(  i
 \sin( -j {\pi b^2 \over 2})
 \right)
  \left(  \cos({\pi b^2 \over 2}) \right)^{-n} \!\! , \ \mathrm{for} \, n\in 2\mathbb{N}
\\[4pt]
 \prod_{j =1 \, mod \, 2}^{n}  \cos( j{ \pi b^2 \over 2} )
\prod_{j =2 \, mod \, 2}^{n-1}
  \left(  i
 \sin( -j {  \pi b^2 \over 2})
 \right)
  \left(  \cos({\pi b^2 \over 2}) \right)^{-n} \! \! , \ \mathrm{for} \, n\in 2\mathbb{N} \!+\! 1 .
\end{array}
\right.
\end{eqnarray*}
It is related to the similar symbol $\qn{.}!$  which we defined in eq.\
(\ref{qsN}) through
\begin{equation}
\qn{n}! = (-1)^{ \frac{1}{12} n(n+1)(2n+1)} (-i)^{n} \, \qnP{n}!\ .
\end{equation}
In order to compare the limiting values (\ref{RW_res}) of Racah-Wigner symbols
(\ref{RW_osp}) with the 6J symbols \eqref{6j-osp} we rewrite the latter in
terms of the new symbol $\qnP{n}$,
\begin{eqnarray}\label{RW_res2}
\nonumber
&&  \hspace{-20pt}
 \left\{ \begin{array}{cccc c}
  -j_1 b && -j_3  b && -j_s b \\
 -j_2 b && -j_4 b  && -j_t b
  \end{array}  \right\}^{\!\! \prime \ \nu_3 \nu_4}_{\nu_1 \nu_2}
  \hspace{-6pt}
= \delta_{\sum_i \nu_i =2(j_s+j_t) \, mod \, 2}
\, {(-1)^{A'(j_i)}  \Delta'_q(j_s, j_2, j_1) \Delta'_q(j_s, j_3, j_4)
\over  2\cos{ ({\pi b^2\over 2} ) } \cos{ ({\pi  \over 2 b^2} ) }  }
     \\[4pt]
   &&
\Delta'_q(j_t, j_3, j_2)\Delta'_q(j_4, j_t, j_1)
\sum_{z\geq 0}  (-1)^{X}
(-1)^{ \frac12 z (z-1) + 2z(j_{1234st}+ j_1j_3 +  j_2 j_4+ j_s j_t  )}
 [z+1]_q'!
\\ && \nonumber
  \Big([z-j_{12s} ]_q'! \, [z-j_{34s} ]_q'!  [ z-j_{14t} ]_q'! \,
 [ z-j_{23t} ]_q'!
[ j_{1234} - z]_q'! \,  [j_{13st} - z]_q'! \, [ j_{24st} - z]_q'! \Big)^{-1}
  \end{eqnarray}
where
\bea
(-1)^{A'(j_i)} &=&
(-1)^{-\frac12 - ( j_{1234st} +1 ) (j_1 j_3 + j_2 j_4 + j_s j_t + 1)
+
\frac12 j_{12s}(j_{12s} -1) + \frac12 j_{s34}(j_{s34} -1) }
 \\
&& 
(-1)^{
 \frac12 j_{23t}(j_{23t} -1) + \frac12 j_{14t}(j_{14t} -1)-  F(j_1,j_2,j_s) -F(j_3,j_4,j_s)- F(j_2,j_3,j_t)- F(j_1,j_4,j_t)} \ ,
\\
&& \hspace{-65pt}
(-1)^{F(j_1,j_2,j_3)} = (-1)^{\frac34 j_{123}(j_{123} +1) + j_1 j_2 j_3 +
j_1 j_2 + j_1 j_3 + j_2 j_3} \ .
\eea
In the case when all  $j_i$ are integer, or equivalently all $a_i = 0$,  the sign  $(-1)^X$  defined in eq.\ (\ref{signX})  and $(-1)^{  2 z(j_{1234st}+ j_1j_3 +  j_2 j_4+ j_s j_t  )} $
 both vanish so that we can relate
 the limit of the Racah-Wigner symbol \eqref{RW_res2} to the \Uqosp\  6J coefficients \eqref{6j-osp},
\begin{eqnarray}
&& \hspace{-40pt}
\left\{ \begin{array}{cc cc c}
  -j_1 b && -j_3  b && -j_s b \\
 -j_2b && -j_4 b  && -j_t b
  \end{array}  \right\}^{\!\! \prime \  \nu_3 \nu_4}_{\nu_1 \nu_2}
  \hspace{-6pt}
= \delta_{\sum_i \nu_i =2(j_s+j_t) \, mod \, 2}
\, {(-1)^{A''(j_i)}
\over  2\cos{ ({\pi b^2\over 2} ) } \cos{ ({\pi  \over 2 b^2} ) }  }
   \left[ \begin{array}{cc  c}
  j_1 & j_2  & j_s \\
 j_3 & j_4  & j_t
  \end{array}  \right]_q
\end{eqnarray}
where
\bea
 & & (-1)^{A''(j_i)} = (-1)^{\frac12 - j_{1234st}  (j_1 j_3 + j_2 j_4 + j_s j_t)
- F(j_1,j_2,j_s)-F(j_3,j_4,j_s)-F(j_2,j_3,j_t)-F(j_1,j_4,j_t) } \ .
\eea
Let us emphasize that in arriving at the expressions \eqref{RW_res} for the
limiting values of the Racah-Wigner symbol, the parameters $j_i$ were allowed
to take either integer ($a_i=0$) or half-integer ($a_i =1$) values. We
have now shown that the limit is proportional to the \Uqosp\  6J
coefficients, provided all arguments $j_i$ are integer. In order to find an interpretation of the limit \eqref{RW_res} in the case of half-integer $j_i$,
we will have to bring in a different idea. It is related to an intriguing
duality between the 6J symbol of \Uqosp\  and \Uqtwo.

As was originally noticed in \cite{Saleur:1989gj}, \cite{Ennes:1997kx}, the
\Uqtwo\ quantum numbers \eqref{x-sl2} with the deformation parameter $q'=i \sqrt{q}$
are related to the \Uqosp\ quantum numbers (\ref{qn_osp}) through,
\begin{equation}
 [x]_{q'} = (-1)^{ {1-x \over 2}}  \qnP{x}\ .
\end{equation}
This equation implies a relation between the quantum factorials,
  \begin{equation}
  \qnP{x}! = (-1)^{x (x-1) \over 4} [x]_{q'}!\ .
  \end{equation}
With its help we can rewrite the \Uqosp\ 6J symbol in terms of the
\Uqtwo\ quantum factorials,
\bea
&& \hspace{-10pt}
\left[ \begin{array}{cc  c}
  j_1 & j_2  & j_s \\
 j_3 & j_4  & j_t
  \end{array}  \right]_{q}
   =
    (-1)^{  \sum_{i=1}^4 {j_i\over 2}(j_i -1) +{ j_s \over 2}(j_s-1) + {j_t \over 2}(j_t-1)  -\frac12 j_{st} j_{1234}- \frac12 j_{13} j_{24}  }
   \\
  &&\hspace{-5pt}
\sum_{z\geq 0} \!
{
 (-1)^{ z + 2z j_{1234st} } [z+1]_{q'}!
   \Delta_{q'}(j_s, j_2, j_1) \Delta_{q'}(j_s, j_3, j_4)\Delta_{q'}(j_t, j_3, j_2)\Delta_{q'}(j_4, j_t, j_1)
   \over
   [z-j_{12s} ]_{q'}! \, [z-j_{34s} ]_{q'}! \, [ z-j_{14t} ]_{q'}!
 [ z-j_{23t} ]_{q'}!
[ j_{1234} - z]_{q'}! \,  [j_{13st} - z]_{q'}! \, [ j_{24st} - z]_{q'}! } \! .
\eea
Due to the condition  $j_{i} \in \mathbb{Z}_{\geq0}$  in the \Uqosp\  6J symbol,
the sign $(-1)^{2z j_{1234st} }$ vanishes and one arrives at the following
relation between the 6J symbols \eqref{6j-osp} and \eqref{6j-sl2}
 \bea
 \left[ \begin{array}{cc  c}
  j_1 & j_2  & j_s \\
 j_3 & j_4  & j_t
  \end{array}  \right]_q
     &=&
    (-1)^{  \sum_{i=1}^4 {j_i\over 2}(j_i -1) +{ j_s \over 2}(j_s-1) + {j_t \over 2}(j_t-1)  -\frac12 j_{st} j_{1234}- \frac12 j_{13} j_{24}  }
      \\&&
    {(-1)^{-j_{12} + j_{34} + 2j_s} \over
     \sqrt{[2j_s+1]_{q'} [ 2j_t +1]_{q'}} }
   \,
   \left( \!\! \begin{array}{cc  c}
  j_1 & j_2  & j_s \\
 j_3 & j_4  & j_t
  \end{array} \!\! \right)_{q'}\ .
 \eea
In a similar way we can relate our limit of Racah-Wigner coefficients
(\ref{RW_res2}) to the 6J symbol of U$_{q'}$(sl(2)) even if some of the
arguments $j_i$ assume (half-)integer values. When written in terms of
$[x]_{q'}$, the Racah-Wigner coefficients \eqref{RW_res2} take the
following form,
\def\m{\!-\!\!}
\begin{eqnarray}\label{RW_res3}
\nonumber
&&  \hspace{-20pt}
 \left\{ \begin{array}{cc cc c}
  -j_1 b && -j_3  b && -j_s b \\
 -j_2 b && -j_4 b  && -j_t b
  \end{array}  \right\}^{\!\! \prime \ \nu_3 \nu_4}_{\nu_1 \nu_2}
  \hspace{-6pt}
= \delta_{\sum_i \nu_i =2(j_s+j_t) \, mod \, 2}
\, {(-1)^{A'''(j_i)}  \Delta_{q'}(j_s, j_2, j_1) \Delta_{q'}(j_s, j_3, j_4)
\over  2\cos{ ({\pi b^2\over 2} ) } \cos{ ({\pi  \over 2 b^2} ) }  }
     \\[4pt]
   &&
\Delta_{q'}(j_t, j_3, j_2)\Delta_{q'}(j_4, j_t, j_1)
\sum_{z\geq 0}  (-1)^{X}
(-1)^{ z+ 2(z+1)(j_1j_3 +  j_2 j_4+ j_s j_t  )}
 [z+1]_{q'}!
\\ && \nonumber
  \Big([z\m j_{12s} ]_{q'}! \, [z\m j_{34s} ]_{q'}!  [ z\m j_{14t} ]_{q'}! \,
 [ z\m j_{23t} ]_{q'}!
[ j_{1234}\m z]_{q'}! \,  [j_{13st} \m z]_{q'}! \, [ j_{24st} \m z]_{q'}! \Big)^{-1}
\end{eqnarray}
where
\bea
&&  \hspace{-10pt}
(-1)^{A'''(j_i)} =
(-1)^{\frac12 - (j_{1234st} +2) (j_1 j_3 + j_2 j_4 + j_s j_t)
- F'(j_1,j_2,j_s)-F'(j_3,j_4,j_s)-F'(j_2,j_3,j_t)-F'(j_1,j_4,j_t) },
\\
&&
\hspace{-10pt}
(-1)^{F'(j_1,j_2,j_3)} = (-1)^{ j_1 j_2 j_3 +\frac12( j_1+ j_2 + j_3 )} \ .
\eea
Using the relations (\ref{nu_s_condition}, \ref{nu_t_condition}) and (\ref{j1234}) one may check that
\begin{equation}
(-1)^{  2j_1j_3 + 2 j_2 j_4+ 2j_s j_t  }
=
 (-1)^{ a_s \nu_1 + a_1 \nu_3 + a_4 \nu_4 +  a_1 a_s + a_2 a_4+  a_s+a_t }\ .
\end{equation}
Since the  parameter $z$ is related to the summation parameter $p$ \eqref{int_general} as
$ z = p + j_{34s}$ and the parity of $p$ is tracked by $\nu' = \nu+ \nu_3 + a_s$, we
may relate the sign under the sum in eq.\ (\ref{RW_res3}) to the sign factor
$(-1)^{X}$ that was defined in eq.\ (\ref{signX}),
\begin{eqnarray}
(-1)^{2(z+1) (  j_1j_3 +  j_2 j_4+ j_s j_t  ) } &=&
 (-1)^{2(\nu+ \nu_3 + a_s + j_{34s}+1) (  j_1j_3 +  j_2 j_4+ j_s j_t  ) }
 \\ \nonumber
 &
 =&  (-1)^{\nu( a_s \nu_1 + a_1 \nu_3 + a_4 \nu_4 +  a_1 a_s + a_2 a_4+  a_s+a_t )} = (-1)^X,
\end{eqnarray}
where we used  eq.\ (\ref{j1234}) to check that $\nu+ \nu_3 + a_s + j_{34s} +1\in 2 \mathbb{N} + 2 (\nu + \nu_3 + \nu_4 + a_s) + \nu $.
Thus the limit (\ref{RW_res3}) is proportional to the 6J symbol of finite
dimensional representations of U$_{q'}$(sl(2)),
\begin{eqnarray}
\label{RW6J_osp1}
\left\{ \begin{array}{cc cc c}
  -j_1 b && -j_3  b && -j_s b \\
 -j_2b && -j_4 b  && -j_t b
  \end{array}  \right\}^{\!\! \prime \  \nu_3 \nu_4}_{\nu_1 \nu_2}
  \hspace{-6pt}
&=& \delta_{\sum_i \nu_i =2j_s+2j_t \, mod \, 2}
\, {(-1)^{A'''(j_i)}
\over  2\cos{ ({\pi b^2\over 2} ) } \cos{ ({\pi  \over 2 b^2} ) }  }
\\ \nonumber
&&
  {(-1)^{-j_{12} + j_{34} + 2j_s} \over
     \sqrt{[2j_s+1]_{q'} [ 2j_t +1]_{q'}} }
   \left( \begin{array}{cc  c}
  j_1 & j_2  & j_s \\
 j_3 & j_4  & j_t
  \end{array}  \right)_{q'}\ .
\end{eqnarray}
This concludes our discussion of the limiting Racah-Wigner coefficients
\eqref{RW_res}. Our analysis has shown that the expression we obtained
from our proposal \eqref{RW_osp} is dual to the 6J symbol for finite
dimensional representations of the quantum universal enveloping algebra
\Uqtwo. By construction the latter satisfy the pentagon equation.
Even though we have not demonstrated that the original symbol
\eqref{RW_osp} solved the pentagon identity for arbitrary values of the
weights $\alpha$, our results provide highly non-trivial evidence in
favor of the proposal. Note in particular that our sign factors were
rather crucial in making things work as soon as some of the parameters
had non-zero label $a_i$, what corresponds to R sector of $N=1$ Liouville 
field theory.

It is actually possible to carry things a bit further. As we noted before, 
the evaluation of the Racah-Wigner symbol \eqref{RW_osp} is possible for 
general degenerate parameters. In that case, the limiting values of the 
Racah-Wigner symbol \eqref{RW_Sgen} can be also related to \Uqtwo\   
6J symbols,
\begin{eqnarray}
&& \hspace{-50pt}
\label{RW6J_osp}
\left\{ \begin{array}{cc cc c}
  -j_1 b - j_1' b^{-1} && -j_3  b- j_3' b^{-1}  && -j_s b- j_s' b^{-1}  \\
 -j_2b - j_2' b^{-1} && -j_4 b- j_4' b^{-1}   && -j_t b- j_t' b^{-1}
  \end{array}  \right\}^{\!\! \prime \ \nu_3 \nu_4}_{\nu_1 \nu_2}
\\ \nonumber
&&  \hspace{80pt}
\sim \delta_{\sum_i \nu_i =2j_s+2j_t \, mod \, 2}
   \left( \begin{array}{cc  c}
  j_1 & j_2  & j_s \\
 j_3 & j_4  & j_t
  \end{array}  \right)_{q'}
  \left( \begin{array}{cc  c}
  j_1' & j_2'  & j_s' \\
 j_3' & j_4'  & j_t'
  \end{array}  \right)_{q''}\ ,
\end{eqnarray}
where the deformation parameters are
$q'^2 = -q =  e^{i \pi  (b^2-1) } $ and $q''^2 =  e^{i \pi (b^{-2}-1) }$.
The above factorization occurs when the sign $ (-1)^X$ defined by eq.\ 
(\ref{signX}) cancels the factor $(-1)^B$ from eq.\ \eqref{-1B} multiplied 
by the sign in eq.\ \eqref{RW_res3} and the corresponding one depending 
on $j'_i$, i.e.\ whenever
\bea
&& \hspace{-20pt}
 (-1)^X (-1)^{B }
(-1)^{2z( j_1 j_3 + j_2 j_4 + j_s j_t) + 2z'( j_1' j_3' + j_2' j_4' + j_s' j_t')}
=1\ . 
\eea
We verified this relation for  degenerate parameters $\alpha_i = - j_i b 
- j_i' b^{-1} $ with $a_i=0$ satisfying $ j_i - j_i' \in 2 \mathbb{Z} $ and 
for arbitrary degenerate parameters with  $a_i=1$.

As in the bosonic case \eqref{Fusion-6Js}, we can relate our result with the 
fusion matrix of supersymmetric minimal models. The degenerate representations 
of NSR algebra are parametrized by a pair of  Kac  labels $(2j ,2j')$, satisfying  
$   j_i +  j_i' \in  \mathbb{Z}_{\geq 0}, \,  j_i - j_i' \in 2 \mathbb{Z} $ in the 
NS sector and $   j_i + j_i' \in  \mathbb{Z}_{\geq 0} + \frac12$ in the R sector. It follows from the coset construction 
$$
 {\text{SMM}}_k = (SU(2)_k \times SU(2)_2)/SU(2)_{k+2}\
$$
of supersymmetric minimal models that the fusion matrix is given in terms of  
two  6J  symbols of \Uqtwo\ with deformation parameters $q_1^2 = \exp(2i \pi/(k+2))$ 
and $ q_2^2 = \exp(2i \pi/(k+ 4))$. Taking into account the symmetry  $q_i 
\leftrightarrow q_i^{-1} $, these values match perfectly those in the 6J 
symbols on the right hand side of eq.\ \eqref{RW6J_osp} if we set 
$b^2 = (k+4)/(k+2)$. 

With all these non-trivial test being performed, we trust that our formula 
\eqref{RW_osp} correctly describes the fusing matrix of $N=1$ Liouville field 
theory for both NS and R sector fields.

\section{Conclusions}

In this work we proposed a formula \eqref{RW_osp} for the Racah-Wigner
symbol of the non-compact quantum universal enveloping algebra \Uqosp.
In order to test our proposal we continued the symbol to a discrete set
of parameters $\alpha = - j b - j' b^{-1}, \ j, j' \in \mathbb{Z}_{\geq0} /2$. 
For integer $j \in\mathbb{N}$ and $j'=0$ we recovered the known 
expressions for Racah-Wigner coefficients of finite dimensional \Uqosp\ 
representations. Half integer values $j$ are not related to the 6J symbols 
of \Uqosp\ but rather to those of \Uqtwo. The relation is furnished by a 
duality which extends the known correspondence between finite dimensional 
representations of \Uqosp\ and integer spin representations of \Uqtwo \ to 
the case of half-integer spins. A related extension was also uncovered 
by Mikhaylov and Witten \cite{Witten}. For cases with  $j' \neq 0$ we
also discussed the expected relation with the fusing matrix of unitary 
superconformal minimal models. There are a number of interesting open 
issues that merit further investigation.

As we stressed before, the Racah-Wigner symbol \eqref{RW_osp} should coincide
with the complete fusing matrix of N=1 Liouville field theory in both the
NS and the R sector \cite{Hadasz:2007wi, Chorazkiewicz:2008es},
\cite{Chorazkiewicz:2011zd}. For NS sector representations a related statement 
was established in \cite{Hadasz:2013bwa}. Of course, it would be interesting 
to incorporate R sector representations into this comparison. Our comments on 
the relation with the fusing matrix of minimal models supports such an 
identification very strongly. Assuming that our Racah-Wigner symbol can be 
reinterpreted as the fusing  matrix in N=1 Liouville theory, our expression 
\eqref{RW_osp}, and special cases thereof, should then also describe various 
operator product coefficients in the bulk and boundary theory, and in 
particular the coefficients of boundary operator product expansion, see e.g.\ 
\cite{Recknagel:book} for a review of the relation.

Recently, it has been observed that the operator product coefficients
of N=1 Liouville field theory with central charge $c = 15/2 + 3(b^2+
b^{-2})$ can be factorized into a products of the coefficients in
ordinary (non-supersymmetric) Liouville field theory and those of an
imaginary (time-like) version thereof
 \cite{Wyllard:2011mn, Belavin:2011sw, Schomerus:2012se, Hadasz:2013dza}.
The central charges of the
latter are given by $c_i = 13 + 6 (b_i^2+b_i^{-2})$ for $i=1,2$ with
\[ b_1^2 = \frac12 (b^2-1) \quad , \quad  b_2^2 = 2 (b^{-2}-1)^{-1} =
         - b_1^{-2} -2\ . \]
This suggest a relation between Racah-Wigner symbols of non-compact  \Uqosp\  for
$q=\exp{i\pi b^2}$ and those of U$_{q_i}$(sl(2)) for the two values
$q_1 =\exp(i\pi b_1^2) = \sqrt {-q}$ and $q_2 = \tilde q_1$. Note
that the latter is obtained from the former by modular transformation.
We see sign of such a relation in the limit of discrete parameters  \eqref{RW6J_osp},
where two 6J symbols for finite dimensional representations of  \Uqtwo\ with $q' = e^{i \pi b_1^2}$ and $q'' = e^{i \pi b_2^{-2}}$ occur.
 We plan to investigate the extension of the duality
between \Uqosp\ and \Uqtwo\ to the continuous self-dual
series of representations in future work. It should also be
linked with a strong-weak coupling duality between the
non-compact OSP(2$|$1)/U(1) cigar-like coset model and
double Liouville theory that was described in
\cite{SC2012talk}.
\smallskip

As we recalled in the introduction, the fusing matrix of $N=1$
Liouville field theory should be a central ingredient in the
construction of a new 3-dimensional topological quantum field
theory, just as Faddeev's quantum dilogarithm \cite{Faddeev:1993rs,
Y.:2005tv}, i.e.\ the building block of the fusing matrix on
Liouville field theory, is used to construct SL(2) Chern-Simons
or quantum Teichmueller theory, see e.g.\ \cite{Kashaev:1996kc,
Kashaev:1998fc,kashaev,Teschner:2003em,Dimofte:2011jd,Andersen:2011bt,
Nidaiev:2013bda}. We will explore these aspects of our work in
a future publication.
\bigskip

\noindent
{\bf Acknowledgements:} We wish to thank Leszek Hadasz, J\"org Teschner,
Grigory Vartanov and Edward Witten for discussions and useful comments. This
work was supported in part by the GRK 1670 ``Mathematics inspired by Quantum
Field and String Theory''. The work of PS was supported by the Kolumb Programme KOL/6/2011-I of FNP, by the NCN grant DEC2011/01/B/ST1/01302 and by the European Social Fund grant POKL.04.01.01-00-054/10-00.

\appendix
\section{Double sine functions}
\label{app}

The double sine function $S_b(x)$ is given in terms of  Barnes' double Gamma function
through
\begin{equation}\label{Sb_def}
S_b(x) = \frac{\Gamma_b(x)}{ \Gamma_b(Q-x)}
\end{equation}
and has poles in positions $x$ such that
\begin{equation}
S_b(x)^{-1}=0 \quad  \iff \quad x = -nb - m b^{-1} \, ,  \qquad
 n,m \in \mathbb{Z}_{\geq 0} \, .
\end{equation}
It satisfies the shift relations
\begin{equation}\label{Sshift}
S_b(x+b^{\pm 1}) = 2 \sin(\pi b^{\pm 1} x) \, S_b(x) \ ,
\end{equation}
which imply that one can evaluate
\begin{eqnarray}\label{sin_qb}
\nonumber
&& S_b(- k b ) = \prod_{j =1}^{k} \left( 2 \sin( -\pi j b^2)  \right)^{-1} \, S_b(0)
=   \left(-2\sin{ (\pi b^2) } \right)^{-k} \,{ S_b(0) \over [ k]! } \ ,
\\[-4pt]
\\[-6pt]  \nonumber
&& S_b(-k b -Q ) =  \left(2\sin{(\pi b^2)} \right)^{-k-1} \left(2\sin{(-\pi b^{-2})} \right)^{-1} \, { S_b(0) \over [ k+1]! } \ ,
\end{eqnarray}
for $ k \in \mathbb{N}$, and more general
\begin{equation}\label{SBxy}
 S_b(-x b -y b^{-1} ) =  \left(2\sin{(\pi b^2)} \right)^{-x}  \left(2\sin{(-\pi b^{-2})} \right)^{-y} \, { (-1)^{xy} S_b(0) \over [ x]! [y]'! } \ ,
\end{equation}
for $x,y \in \mathbb{Z}_{\geq0}$.
 We have also used the q-number $[x] = \frac{\sin(\pi b^2 x)}  {\sin{\pi b^2}}$ and $[y]' = \frac{\sin(\pi b^{-2} y)}  {\sin{\pi b^{-2}}}$.

The supersymmetric double sine functions are constructed from Barnes'
double Gamma functions
\begin{eqnarray}\label{sSine}
\nonumber
S_1(x) &=& S_{NS}(x) = {
 \Gamma_b\left( \frac{x}{2} \right) \,   \Gamma_b\left( \frac{x+Q}{2} \right)
\over
 \Gamma_b\left( \frac{Q-x}{2} \right) \Gamma_b\left( \frac{2Q-x}{2} \right)
}
\\[-4pt]
\\[-4pt] \nonumber
S_0(x) &=& S_{R}(x) = {
 \Gamma_b\left( \frac{x + b}{2} \right) \,   \Gamma_b\left( \frac{x + b^{-1}}{2} \right)
\over
 \Gamma_b\left( \frac{Q-x +b }{2} \right) \Gamma_b\left( \frac{Q-x + b^{-1} }{2} \right)
}
\end{eqnarray}
and they have poles as
\begin{eqnarray} \label{Snu_poles}
 S_\nu(x)^{-1} =0  \quad \iff \quad x = k b + l/b\, , \quad k,l \in \mathbb{Z}_{\geq0}\, ,
 \quad  k+l \in 2 \mathbb{N} - 1 - \nu.
\end{eqnarray}
They obey the shift relations:
\begin{equation}\label{shiftR}
S_1(x+ b^{\pm 1}) = 2 \cos( {\pi b^{\pm 1} x \over2} ) S_0(x), \qquad
S_0(x+ b^{\pm 1} ) = 2 \sin( {\pi b^{\pm 1}  x \over2} ) S_1(x).
\end{equation}
For $ x$ integer such that $x \in 2\mathbb{N} - 1-\nu $  the double sine functions can be written as:
\begin{eqnarray} \label{Sx}
&&
\hspace{-15pt}
S_\nu(-x b ) =
{  S_1(0) \over \left(2\cos{ ({\pi b^2\over 2} ) } \right)^{x}  \qn{x}! }
\\ \nonumber
&&
\hspace{-15pt}
 S_\nu(-x b \! -Q ) \!=\!
 {  (-1)^{-\frac{x+1}{2} - \frac12 \delta_{\nu,1}} \,  S_1(0) \over
 2\cos{ ({\pi\over 2 b^2} ) } \! \left(2\cos{ ({\pi b^2\over 2} ) } \right)^{x+1} \!
  \qn{x+1}! }
 =\!
   { (-1)^{-\frac{x(x-1)}{2}+1} \,  S_1(0)
   \over
    2\cos{ ({\pi\over 2 b^2} ) } \!  \left(2\cos{ ({\pi b^2\over 2} ) } \right)^{x+1} \!
    \qn{x+1}! }
\end{eqnarray}
where
\begin{eqnarray}
\label{[n]q}
\qn{n}! =
\left\{
\begin{array}{l}
\prod_{j =1 \, mod \, 2}^{n-1}  \cos(j { \pi b^2 \over 2})
\prod_{j =2 \, mod \, 2}^{n}
 \sin( -j {\pi b^2 \over 2})
 \left(  \cos({\pi b^2 \over 2}) \right)^{-n} , \ \mathrm{for} \, n\in 2\mathbb{N}
\\[4pt]
 \prod_{j =1 \, mod \, 2}^{n}  \cos( j{ \pi b^2 \over 2} )
\prod_{j =2 \, mod \, 2}^{n-1}
 \sin( -j {  \pi b^2 \over 2})
  \left(  \cos({\pi b^2 \over 2}) \right)^{-n} , \ \mathrm{for} \, n\in 2\mathbb{N} +1.
\end{array}
\right.
\end{eqnarray}
In general, for arguments  such that $ x+y \in 2 \mathbb{N} - 1 - \nu$, the double sine functions satisfy the  identity:
\begin{equation}\label{SnuXY}
 S_\nu(-x b -yb^{-1} ) =
   { (-1)^{\frac{xy}{2} +  \nu \frac{x^2}{2} } \,  S_1(0)
   \over
 \left(2\cos{ ({\pi b^2\over 2} ) } \right)^{x}  \left(  2\cos{ ({\pi\over 2 b^2} ) } \right)^{y} \,
    \qbp{x}! \qbm{y}! }
\end{equation}
where $\qbm{n}!$ is given by the formula \eqref{[n]q} with $b$ exchanged for $b^{-1}$.

\end{document}